\documentclass[APA,Times2COL]{WileyNJDv5} 
\usepackage{listings}
\usepackage{tabularx}

\articletype{Industrial Application}%

\received{Date Month Year}
\revised{Date Month Year}
\accepted{Date Month Year}
\journal{Journal}
\volume{00}
\copyyear{2023}
\startpage{1}

\usepackage{soul}
\sethlcolor{yellow}
\begin{document}

\title{Configurable convolutional neural networks for real-time pedestrian-level wind prediction in urban environments}

\author[1]{Alfredo Vicente Clemente}

\author[2,3]{Knut Erik Teigen Giljarhus}

\author[3]{Luca Oggiano}

\author[1,4]{Massimiliano Ruocco}

\authormark{CLEMENTE \textsc{et al.}}
\titlemark{Configurable convolutional neural networks for real-time pedestrian-level wind prediction in urban environments}

\address[1]{\orgdiv{Software Engineering, Safety and Security}, \orgname{SINTEF Digital}, \orgaddress{\state{Trondheim}, \country{Norway}}}

\address[2]{\orgdiv{Department of Mechanical and Structural Engineering and
Materials Science}, \orgname{University of Stavanger}, \orgaddress{\state{Stavanger}, \country{Norway}}}

\address[3]{\orgname{Nablaflow AS}, \orgaddress{\state{Stavanger}, \country{Norway}}}

\address[4]{\orgdiv{Department of Computer Science}, \orgname{Norwegian University of Science and Technology}, \orgaddress{\state{Trondheim}, \country{Norway}}}

\corres{Corresponding author Knut Erik T. Giljarhus. \email{knut.e.giljarhus@uis.no}}



\abstract[Abstract]{

Urbanization has underscored the importance of understanding the pedestrian wind environment in urban and architectural design contexts. Pedestrian Wind Comfort (PWC) focuses on the effects of wind on the safety and comfort of pedestrians and cyclists, given the influence of urban structures on the local microclimate. Traditional Computational Fluid Dynamics (CFD) methods used for PWC analysis have limitations in computation, cost, and time. 
Deep-learning models have the potential to significantly speed up this process. The prevailing state-of-the-art methodologies largely rely on GAN-based models, such as pix2pix, which have exhibited training instability issues. In contrast, our work introduces a convolutional neural network (CNN) approach based on the U-Net architecture, offering a more stable and streamlined solution. 
The process of generating a wind flow prediction at pedestrian level
is reformulated from a 3D CFD simulation into a 
2D image-to-image translation task, 
using the projected building heights as input.
Testing on standard consumer hardware shows that our model can efficiently predict wind velocities in urban settings in real time. Further tests on different configurations of the model, combined with a Pareto front analysis, helped identify the trade-off between accuracy and computational efficiency. This CNN-based approach provides a fast and efficient method for PWC analysis, potentially aiding in more efficient urban design processes.}



\maketitle


\renewcommand\thefootnote{\fnsymbol{footnote}}
\setcounter{footnote}{1}

\section{Introduction}\label{introduction}
Urbanization presents an increasing need for efficient pedestrian wind environment analysis in urban and architectural design. Traditional techniques such as computational fluid dynamics (CFD) are known to be computationally heavy, expensive, and time-consuming. In this paper, we propose a deep learning-based tool as an alternative model, with the potential to enhance the efficiency and affordability of this analysis process. 

Pedestrian Wind Comfort (PWC) is a specialized area within wind engineering that studies how wind affects the comfort and safety of pedestrians and cyclists. This field is critical because building construction can significantly alter the local microclimate, creating potentially uncomfortable or dangerous conditions. For instance, wind interacting with large structures can cause downdraft and wind acceleration, leading to discomfort for pedestrians, and even posing safety risks. As such, wind comfort studies, often facilitated by tools like CFD, have become integral to urban planning and civil engineering, helping to predict and mitigate potential wind comfort issues before construction, ensuring the safety and comfort of urban inhabitants and visitors. 

PWC is calculated by combining the wind statistics for a given location with the simulated wind velocities of the location from multiple different wind directions.  
PWC is utilized as part of modern building design to ensure that the effect of new constructions on the wind microclimate is limited. PWC is calculated several times throughout the design process, and is used to inform design choices. Current solutions for developing PWC require days or weeks of work and require specialized equipment and knowledge, severely limiting the iteration speed of the design process. A real-time PWC system would remove both of these limitations, significantly reducing iteration time during the design process of a modern building.

CFD has matured as a technology to predict the
wind velocities at pedestrian level\cite{blocken2004pedestrian,blocken201450}.
There are also robust guidelines available for numerical settings and turbulence models\cite{franke2004recommendations,tominaga2008aij}.
Recent studies have shown that although the bluff bodies present in an urban environment cause vortex shedding and highly transient flow, they can be reasonably predicted with steady-state turbulence modelling\cite{blocken2018over,tominaga2023accuracy}.
Increased computational power and more robust meshing technologies and computational methods have made it possible to routinely simulate large urban environments. Although there are efforts to create highly specialized solvers running at higher speeds\cite{WANG2021107586,mortezazadeh2020solving,dai2022evaluation},
the time of a single simulation is typically
measured in hours, even running on high-performance computing hardware. For pedestrian comfort, multiple simulations from different wind directions are also needed, further increasing the simulation time. This motivates the use of data-driven techniques to speed up the solution.

In recent research, machine learning techniques, particularly deep learning (DL) methods, have been increasingly employed to investigate wind characteristics and their influence on buildings and urban environments. One of the first works\cite{kim2021predicting} in this direction presents a noteworthy application of DL techniques in the context of urban wind analysis. In this study, DL methodologies were employed to address a specific challenge: interpolating discrete wind environment measurement points around buildings. The primary goal was to achieve accurate predictions of wind speed values in areas where measurements were not available, thereby obtaining a comprehensive wind field representation. 

Some advancement in this field is the utilization of convolutional neural networks (CNNs)\cite{convnet}. CNNs have demonstrated their efficacy in design exploration tasks by capturing complex, non-linear relationships between input and output data and by extracting spatial relationships. For example, a pioneering work used CNNs to predict velocity distributions around a circular cylinder within a flow field\cite{jin2018prediction}. Similarly, another research work\cite{tanaka2019optimization} trained a CNN to predict wind distribution around a building complex. Their training dataset was derived from a self-developed urban generation tool, coupled with batch CFD simulations.

In our work, we redefine the problem from its conventional form, where 3D flow fields are computed through computational fluid dynamics, to a novel approach. Specifically, we frame the task as a 2D image-to-image translation problem \cite{isola2017image} centered on building footprints similarly to other works \cite{mokhtar2020, mokhtar2021, hoeiness2021} . The image-to-image translation problem in AI revolves around the task of converting an input image from one domain or style into an output image in a different domain or style while preserving the semantic content of the input. In computer vision this technique has been used for example for image synthesis \cite{park2019semantic} image segmentation \cite{guo2020gan}, style transfer \cite{pathak2016context} and image super resolution \cite{yuan2018unsupervised}. In this approach, we represent the source domain with pixels signifying building heights, and in the target domain, these pixels represent wind speeds in a specific direction. The core objective of our methodology is to facilitate this transformation in an efficient manner.

Generative adversarial networks (GANs) have emerged as a pivotal deep learning methodology \cite{gans}, particularly in the domain of computer vision. Their ability to generate data similar to the input data has led to their widespread adoption and significant advancements in various applications. A notable variant of GANs is the conditional GAN, exemplified by the pix2pix model \cite{isola2017image}. Pix2pix operates on the principle of image-to-image translation, where it aims to map an input image to an output on a pixel-by-pixel basis. In the context of building environments, pix2pix has been instrumental in different works \cite{mokhtar2020, mokhtar2021pedestrian, duering2020optimizing, hoeiness2021}. The works of \cite{mokhtar2020, mokhtar2021pedestrian} utilized pix2pix to understand wind distribution around architectural structures. Their approach was underpinned by training sets derived from batch computational fluid dynamics (CFD) simulation. \citeauthor{duering2020optimizing}\citeyear{duering2020optimizing}  integrated a pix2pix-based model for outdoor wind flow prediction with a solar radiation prediction system. Their integrated approach aimed to streamline urban design and optimization processes, addressing challenges related to interoperability and computational overheads. Finally, \citeauthor{hoeiness2021}\citeyear{hoeiness2021} compared the performance of pix2pix with other methods to assess the wind prediction in urban areas by introducing attention mechanisms and spectral normalization to facilitate stable training. Other examples of use of conditional GAN (not pix2pix) are related to FlowGAN\cite{Chen2020} and Boundary Condition CGAN, BC-GAN\cite{faulkner2023fast}. FlowGAN emerged as a solution to the challenges of resource-intensive computational fluid dynamics (CFD) simulations in flow-related design optimization. As a conditional GAN, FlowGAN predicts flow fields under various conditions without re-training, demonstrating reduced prediction errors and superior generalization compared to other methods. BC-GAN is used to predict indoor airflow distribution. This model generates 2D airflow distribution images based on continuous input parameters and has shown to predict velocity and temperature distributions with good performance. 




Image-to-image methods are often solved using an U-Net architecture\cite{unet}. When it comes to image-to-image translation tasks, both GAN-based architecture and U-Net can be employed, but U-Net has certain advantages over GANs. For example, the simplicity and directness. U-Net is a straightforward encoder-decoder architecture without the adversarial components present in GANs. This makes U-Net easier to train and less prone to mode collapse or other training instabilities that can sometimes plague GANs. Training a GAN involves a two-player minimax game between the generator and discriminator, which can be tricky to stabilize. U-Net does not have this complexity, leading to a more straightforward training process. Since the development of original U-Net, several improvements have been made to make CNNs significantly more efficient and performant. The Half-U-Net \cite{half_unet} simplifies the decoder, reducing the amount of parameters and required computation, while increasing performance. ConvNext \cite{convnext} provide several improvements over traditional CNNs significantly improving performance.

The main contributions of our work are as follows:
\begin{itemize}
    \item We reformulate the wind flow prediction at pedestrian level
     from a 3D CFD simulation into a 2D image-to-image translation task, 
using the projected building heights as input.
    \item We introduce a novel, configurable Convolutional Neural Network (CNN) model. This model is adept at swiftly generating precise wind velocities in urban settings, achieving high accuracy within milliseconds on standard consumer hardware.
    \item We conduct tests on various configurations of our proposed method, utilizing a Pareto front analysis. This aids in identifying the optimal balance between loss and runtime.
\end{itemize}


\section{Dataset}\label{dataset}
\begin{figure*}
\includegraphics[width=0.5\textwidth]{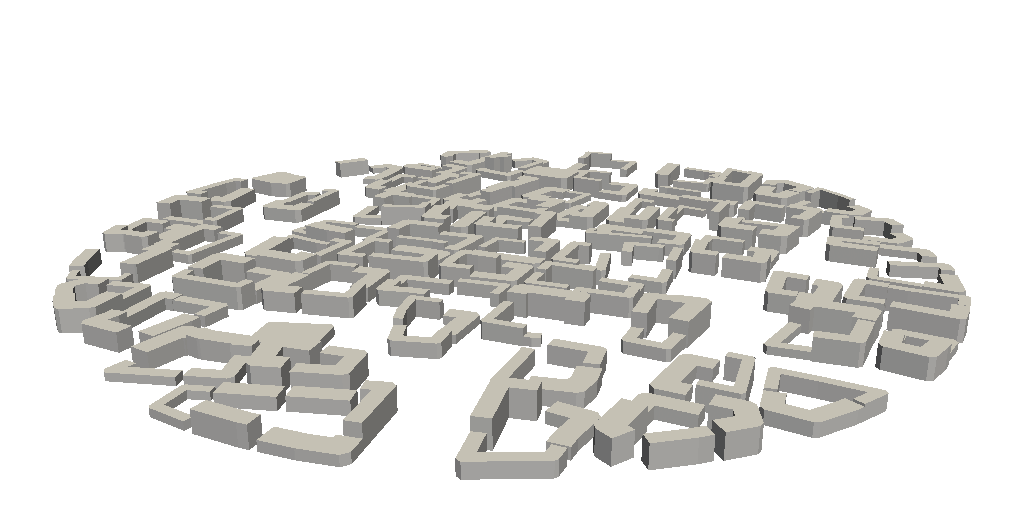}
\includegraphics[width=0.5\textwidth]{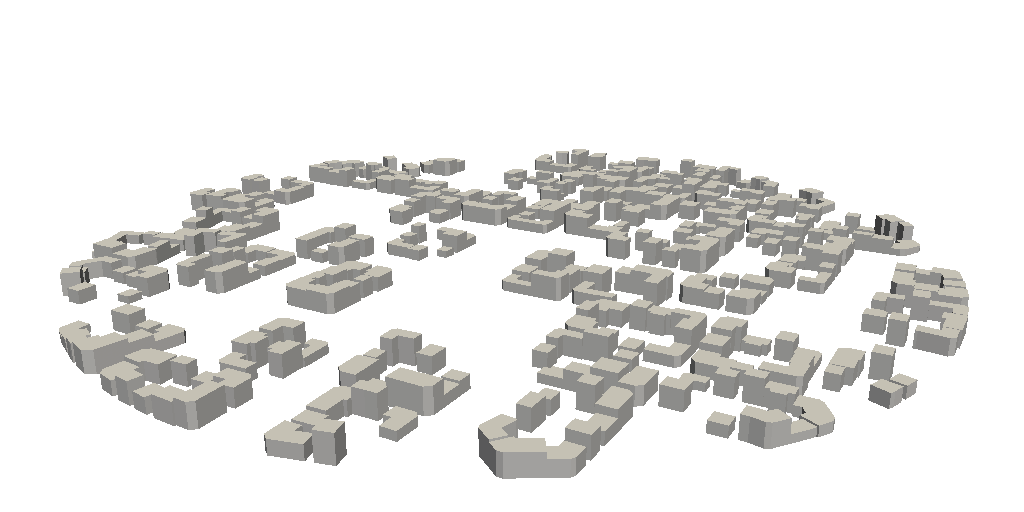}
\includegraphics[width=0.5\textwidth]{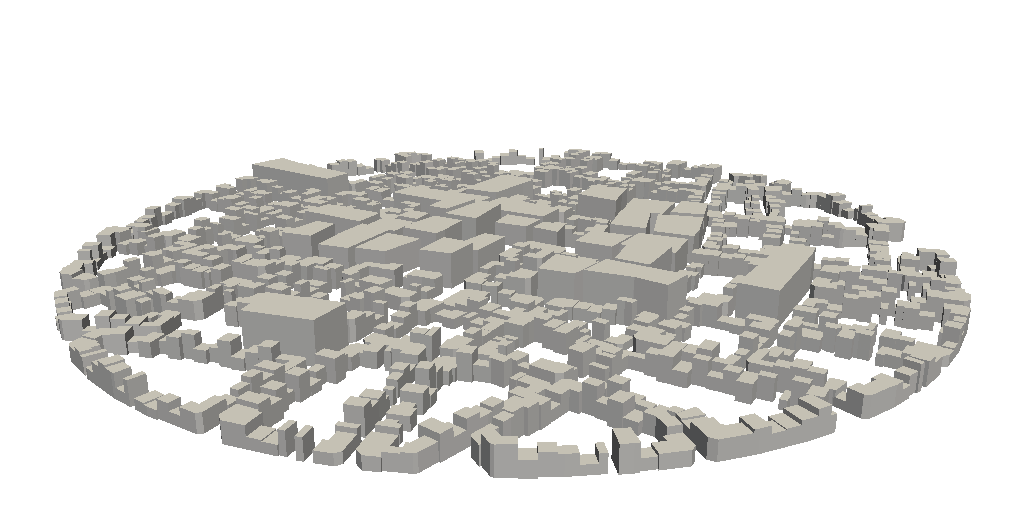}
\includegraphics[width=0.5\textwidth]{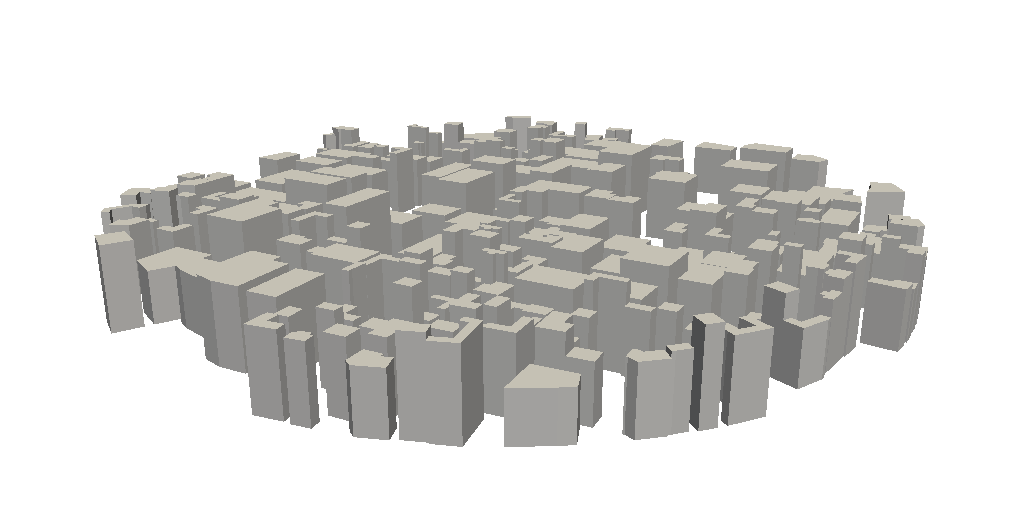}
\includegraphics[width=0.5\textwidth]{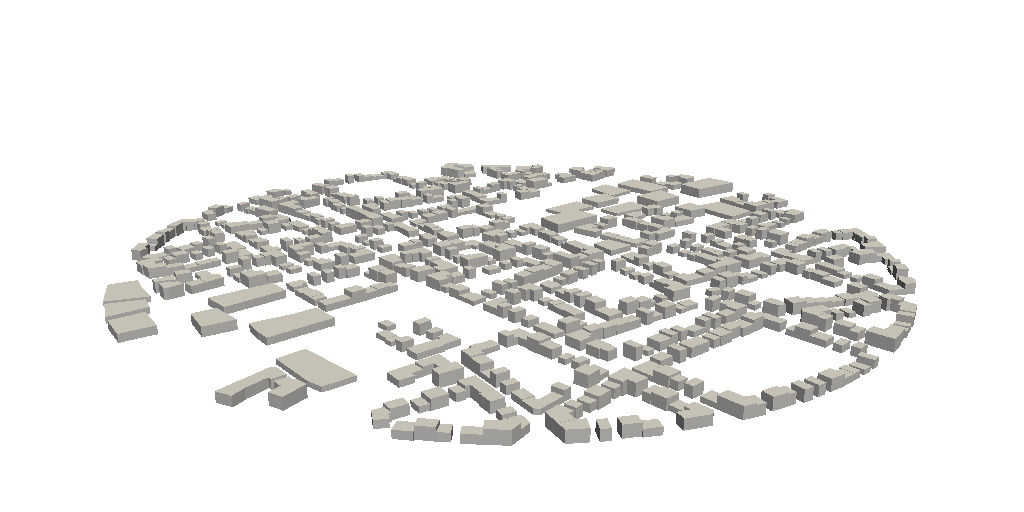}
\includegraphics[width=0.5\textwidth]{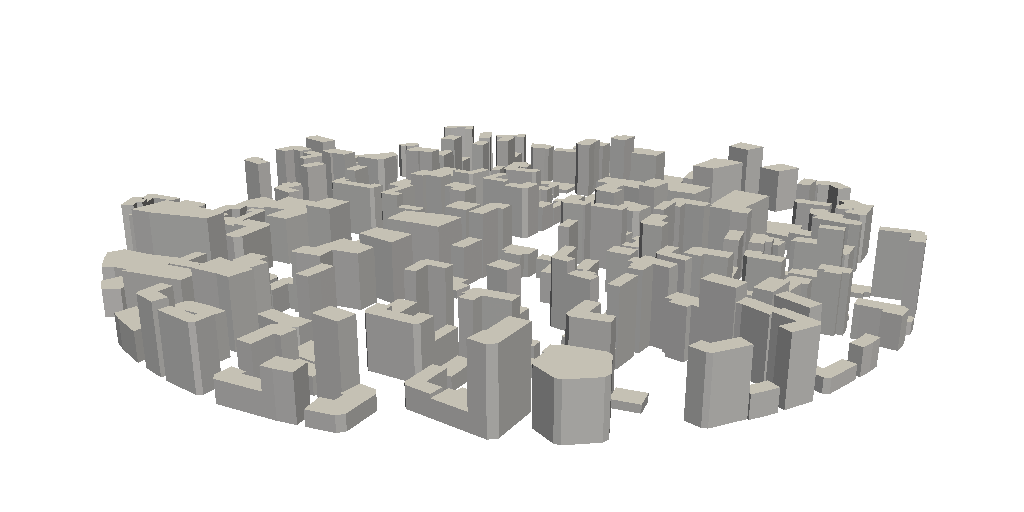}
\caption{Examples of urban city geometries from the dataset.}
\label{fig:geom}
\end{figure*}

The dataset consisted of 163 generated urban scenes. 
The dataset was built in collaboration with an architect to achieve a broad and realistic representation of urban city morphologies. Figure~\ref{fig:geom} shows six examples from the
dataset, illustrating how the data includes different building shapes,
building heights and urban densities. We note that the emphasis is on simple building shapes, typical of an early-design process. A circular border of buildings was used to ensure that there was no bias with wind coming from
different wind directions. The total diameter of the building region was 1100 m, and the building heights ranged from
6 m to 100 m.

A CFD simulation is performed on each urban scene, with
wind coming from 8 different wind directions. This results in a total of 1304 discrete sets of data in the dataset.
The CFD simulations are performed using the open-source CFD
simulation OpenFOAM, v2206. 
The flow is governed by the Reynolds-averaged
Navier-Stokes (RANS) equations,
\begin{align}
    \nabla \cdot \mathbf u &= 0 \\
    \nabla\cdot(\mathbf u \mathbf u) &= -\nabla p + \nabla\cdot(\nu_\text{eff}\nabla\mathbf u),
\end{align}
where $\mathbf u$ is the velocity, $p$ is the modified pressure (divided by density),
and $\nu_\text{eff}$ is the effective kinematic viscosity. 
The turbulence is modeled by adding an additional viscosity
on top of the fluid viscosity. This viscosity is calculated
from two additional fields representing turbulent
kinetic energy, $k$, and turbulent dissipation rate, $\epsilon$,
\begin{equation}
    \nu_\text{eff} = \nu + \nu_t = \nu + C_\mu\frac{k^2}{\epsilon}.
\end{equation}
The two additional fields are found from solving additional partial differential equations. In this work, the realizable version of the standard $k$-$\epsilon$ model is used.

As the RANS equations are non-linear and weakly coupled,
they require an iterative procedure to solve. Here, the
SIMPLE algorithm was used for this purpose. Second-order finite
volume discretization schemes were used for each term in the equations.

\begin{figure}[tbp]
\centering
\includegraphics[width=0.48\textwidth]{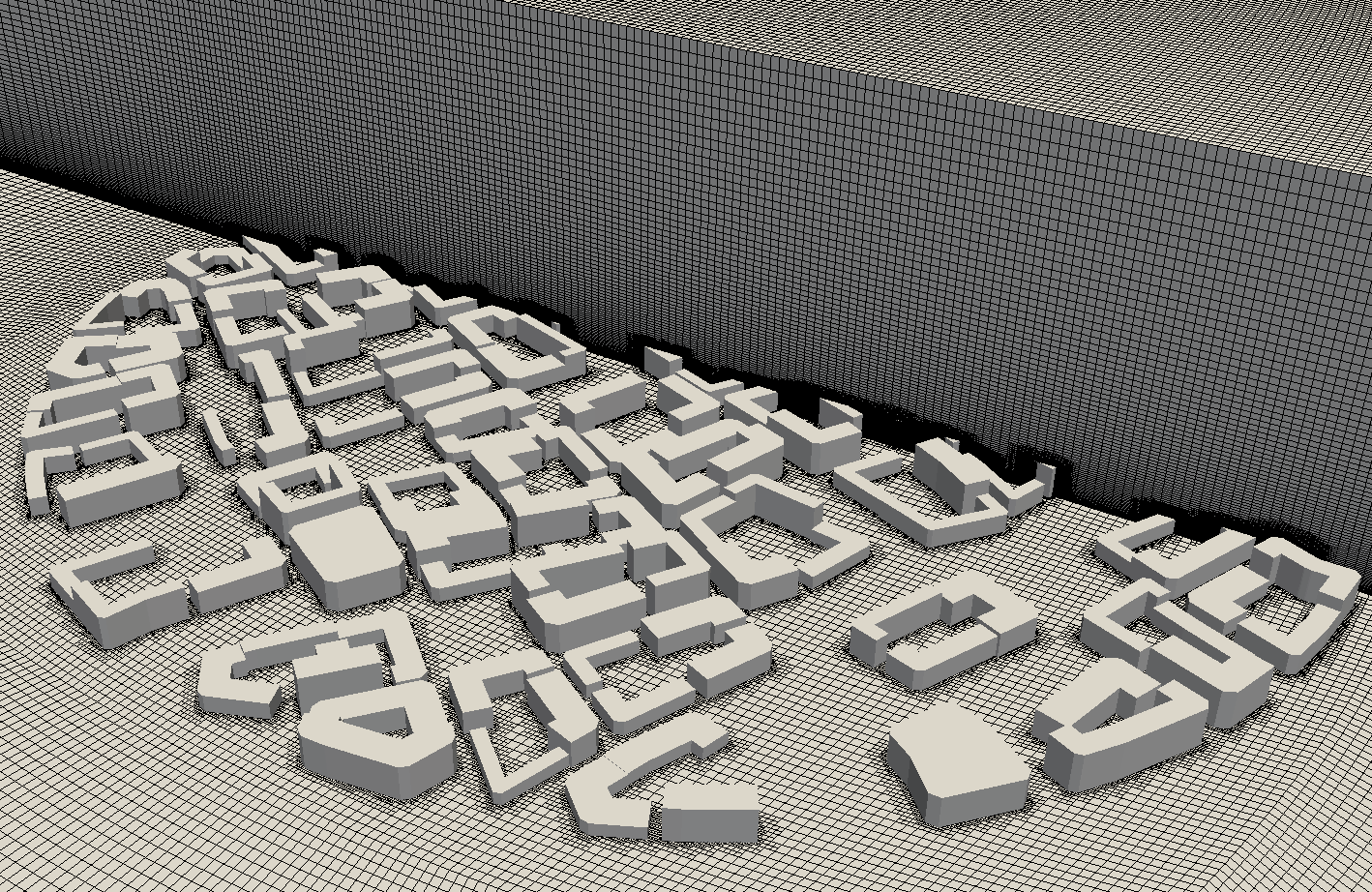}
\caption{Example of computational mesh used for 
CFD simulation.}
\label{fig:mesh}
\end{figure}

The computational mesh was generated using the snappyHexMesh, a hex-dominant,
unstructured mesh generator. Two levels of 1:2 refinement were used near the buildings, giving a cell size near the buildings of 1.5 m. The cells were also stretched towards the ground so that the height of the first grid cell is 0.2 m and there were 10 grid cells below the 2 m level, where the training data
is extracted.
In total, the number of grid cells were around 10 million cells. An image of the computational mesh for one of the cases is shown in Figure \ref{fig:mesh}.
The total computational domain had a diameter of 3000 m and a height of 300 m. At the inlet, a logarithmic atmospheric boundary layer profile is imposed, having a reference velocity measuring 5 m/s at a 10 m elevation from the ground. The terrain had a rough wall characterized by a 0.03 m roughness length, while standard no-slip conditions were used for the buildings. 
The computational setup is similar to recent work by the authors, which also include mesh sensitivity studies and validation examples\cite{haagbo2021influence,haagbo2022pedestrian}.
We ran the simulations on the Amazon AWS infrastructure, through the ArchiWind web platform\cite{archiwind}.
The c5d.metal instances were used, which consists of Intel Xeon Platinum 8275CL processors with 96 3.0 GHz CPU cores.
Each simulation is run in serial, as this avoids the overhead of communication between the cores.
The total simulation time for a single case was around 4 h, running on a single CPU. 

For input to the ML model, each geometry is represented with 25 images of size $2048 \times 2048$ pixels, where one image represents the height-map of the buildings, and the remaining images represent the simulated wind speeds at a height of 2 m above the ground from 8 different directions in the $x$, $y$ and $z$ axis. Training is performed on 80\% of the urban scenes, 10\% are used for validation, and the final 10\% are used for testing. Images are down-scaled to $1024 \times 1024$ pixels before being ingested by the model.
The conversion of velocities from the simulation grid to an
image means an interpolation from an unstructured grid to a
structured grid with fixed cell sizes. Also, the velocities
are converted into a discrete integer value between 0 and 255, causing some loss of information.

\section{Methods}\label{methods}

\textbf{Problem definition.} 

We condense the 3D geometry of the buildings into a 2D representation, using a projected map of the building height. This approach transforms the task of predicting the wind field into an image-to-image translation task.

\begin{figure}[h!]
    \centering
    \includegraphics[width=.40\textwidth]{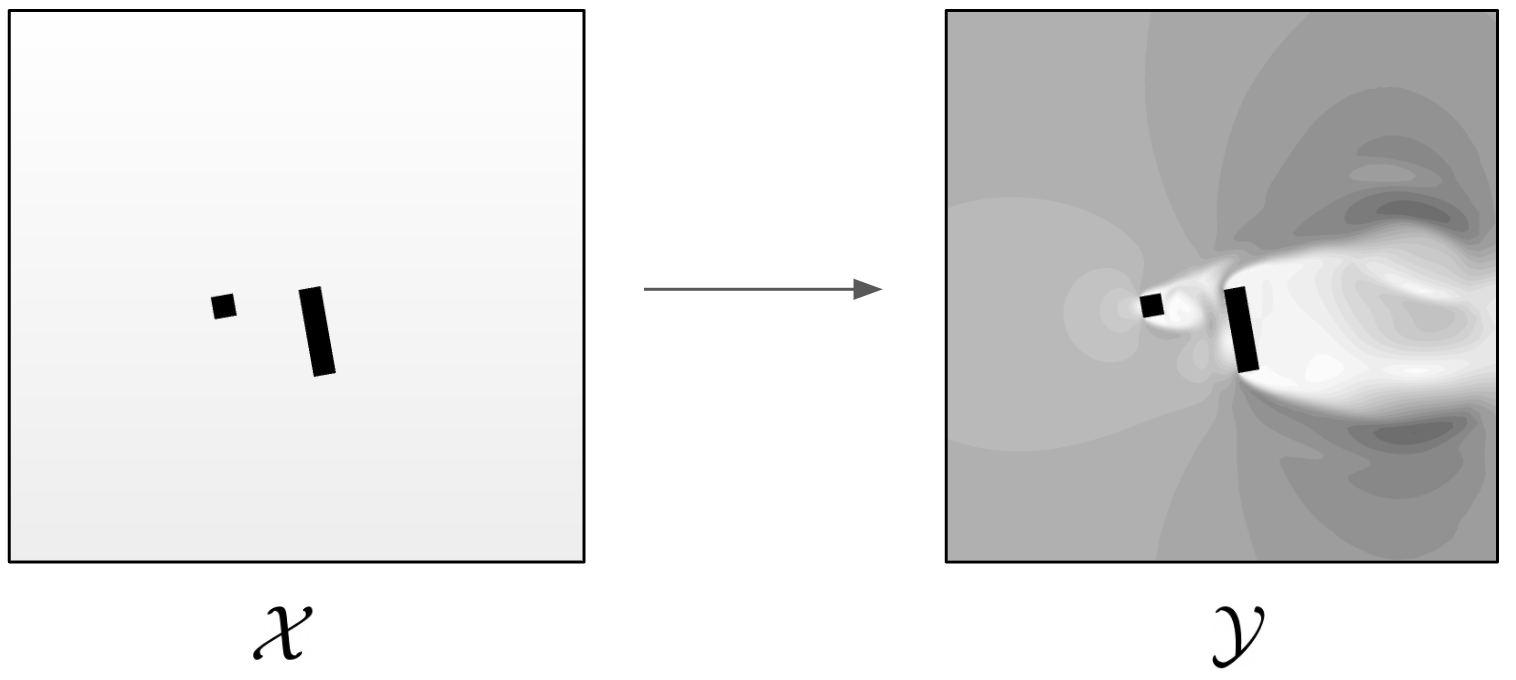}
    \caption{Mapping from domain $\mathcal{X}$ to $\mathcal{Y}$.}
    \label{fig:domain-mapping}
\end{figure}

We consider two distinct domains: $\mathcal{X}$, representing the 2D height map, and $\mathcal{Y}$, denoting the wind velocity prediction. Our objective is to ascertain the functions that map these domains, 
$\mathcal{X} \rightarrow \mathcal{Y}$, between the image pair $(x_i, y_i)$, with $x_i \in \mathcal{X}$ and $y_i \in \mathcal{Y}$, as shown in \autoref{fig:domain-mapping}. 
The system is designed to accept the 2D height map as a condition for the wind flow prediction, i.e. $G: x \rightarrow y$,
such that the mapping occurs between the image pairs
and not just between the two domains.

\begin{figure}[ht]
\centerline{\includegraphics[width=0.5\columnwidth]{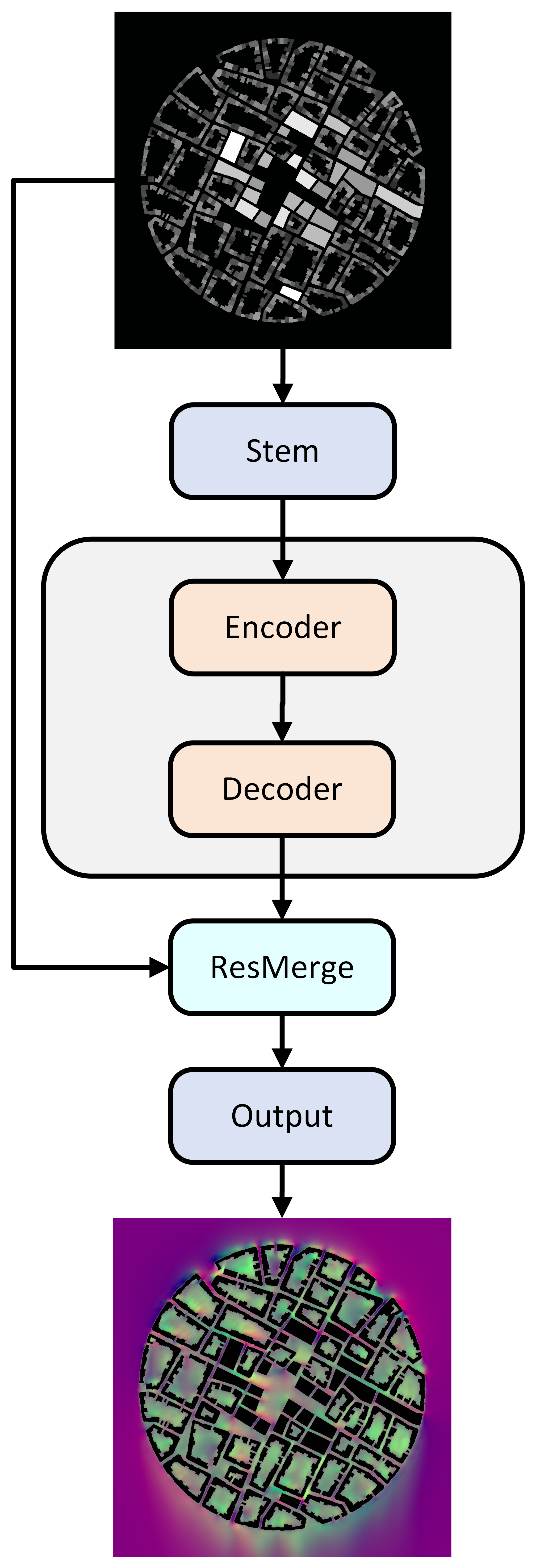}}
\caption{Network architecture consisting of a stem, encoder, decoder, ResMerge block and output block.\label{fig:architecture}}
\end{figure}

\textbf{Proposed methods.} We propose a configurable convolutional neural network architecture shown in Figure \ref{fig:architecture}. This architecture is structured around four primary blocks: the stem, the encoder-decoder, the ResMerge, and the output block. 

\begin{figure}[ht]
\centerline{\includegraphics[scale=0.20]{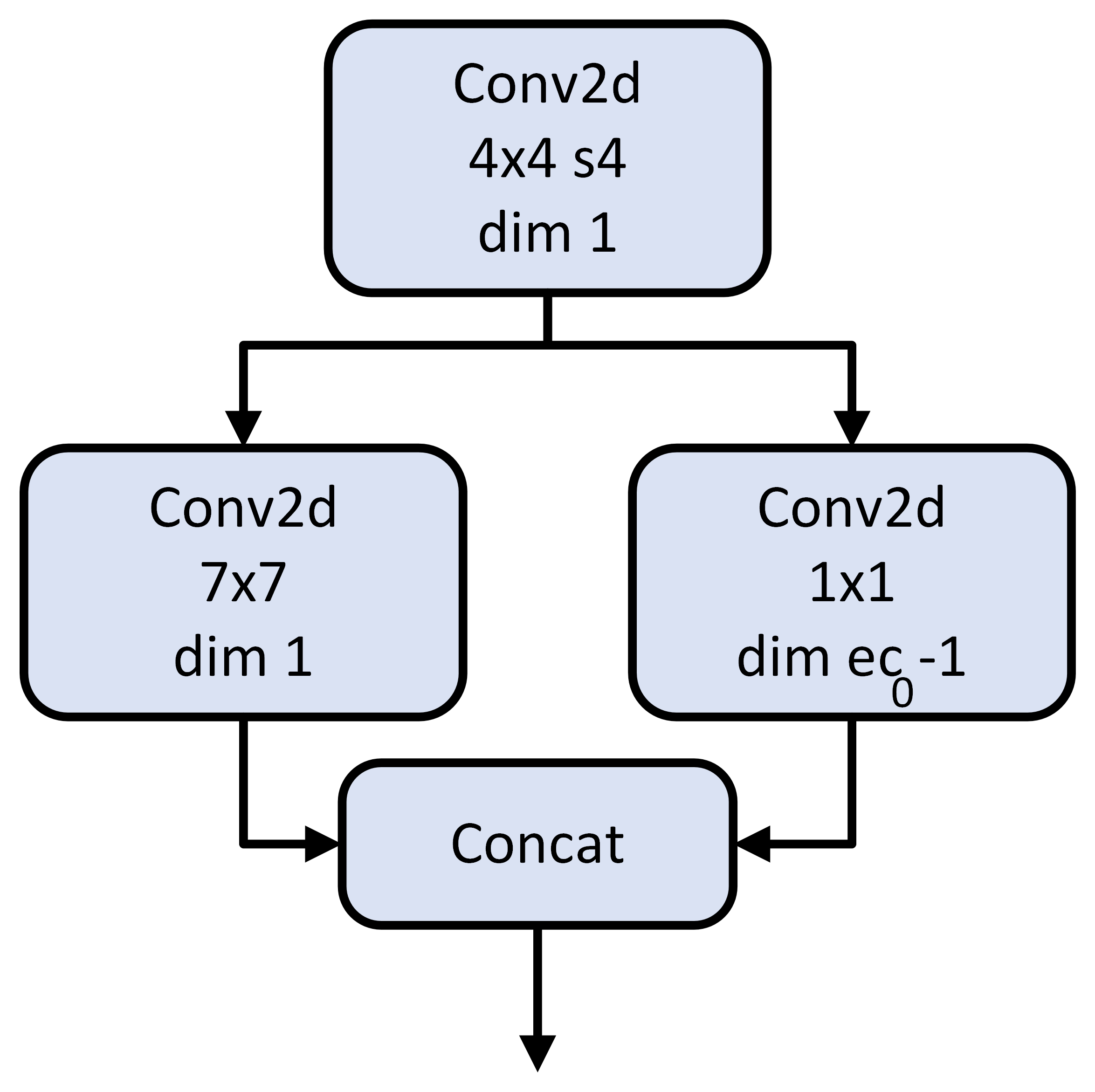}}
\caption{The stem block has a 4x4 convolutional layer that branches into 7x7 and 1x1 layers, with their outputs concatenated channel-wise.\label{fig:stem}}
\end{figure}

First, the \textit{stem block} serves as the initial layer of our architecture. It incorporates a 2D convolutional layer with a kernel size of 4x4. Notably, the output dimension of this layer is designed to match the input dimension, which, for the scope of our experiments, remains consistent at one for all the experiments. The output from this layer is concurrently channeled into two distinct 2D convolutional layers: the first layer has a kernel size of 7x7, with its output dimension mirroring the input dimensions while the second layer features a kernel size of 1x1. Its output dimension is determined by subtracting the input dimension from the first encoder channel's size. The outputs from these two layers are then concatenated along the channel axis, forming a unified output. The design of this stem module is consistent across all configurations, as illustrated in Figure \ref{fig:stem}, and is inspired by the guidelines of ConvNext\cite{convnext}.

\begin{figure}[ht]
\centerline{\includegraphics[scale=0.20]{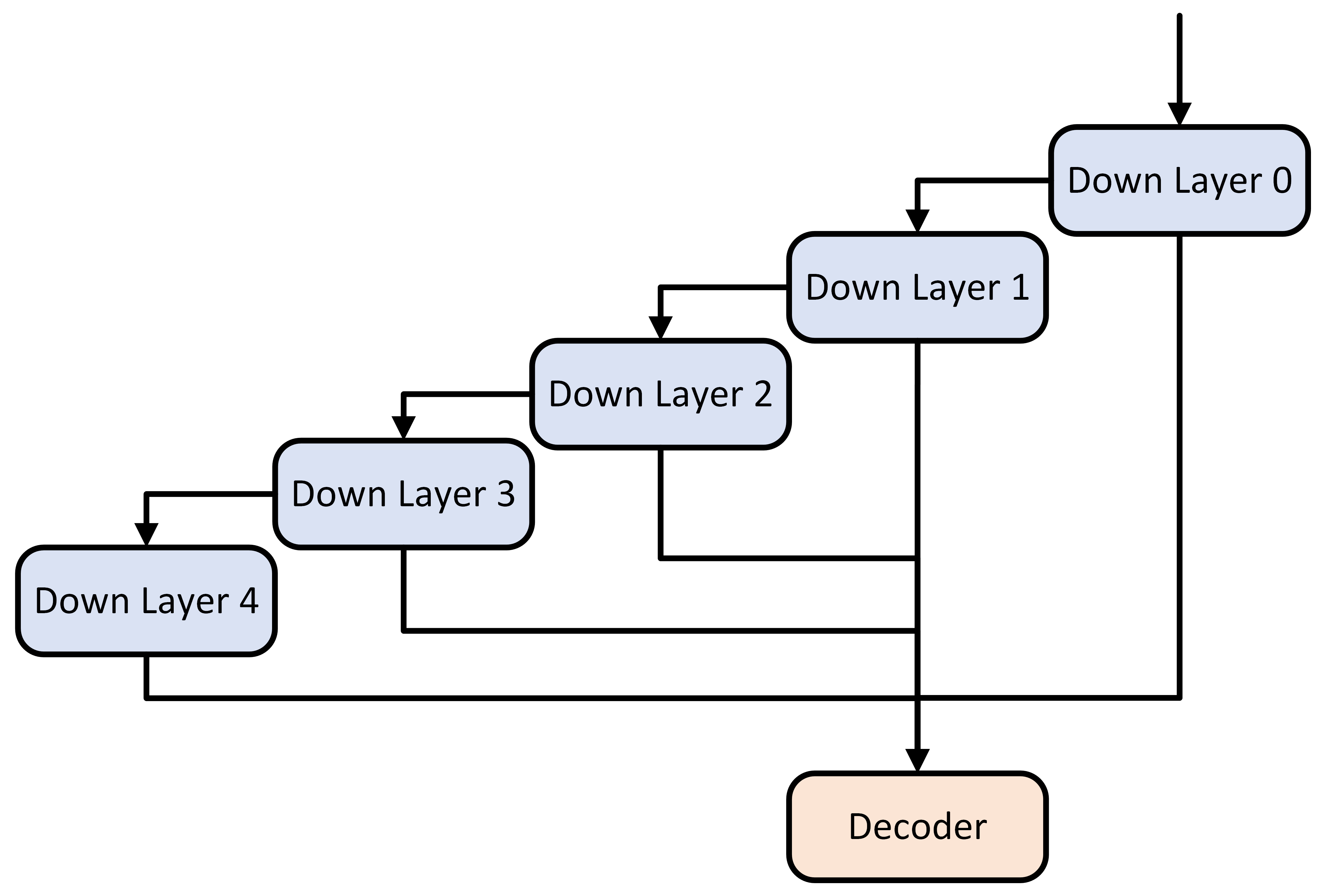}}
\caption{U-Net encoder.\label{fig:unet_encoder}}
\end{figure}

After the stem block, the \textit{encoder-decoder block} follows as shown in Figure \ref{fig:unet_encoder}. The encoder is based on the U-Net\cite{unet} architecture, which
was originally developed for biomedical image segmentation, and adopts an auto-encoder framework. 
The architecture of this framework includes an encoder, which compresses the input via convolutional layers and max-pooling methods. Subsequently, a decoder is employed to enlarge the condensed output through the application of up-sampling techniques.
To retain high-resolution features from the input, the encoder's features are merged with those of the decoder during the up-sampling process. This merging, termed skip-connections, involves concatenating channels from one layer with another. Consequently, the decoder mirrors the encoder's structure, leading to a U-shaped architecture, which is the origin of the name "U-Net". This design, with its skip connections, facilitates efficient information flow across layers, especially through the bottleneck between the encoder and decoder. In image-to-image translation tasks, U-Net with skip-connections is advantageous as it ensures efficient information transfer across layers, particularly bridging the bottleneck between the encoder and decoder. 

\begin{figure}[ht]
\centerline{\includegraphics[scale=0.20]{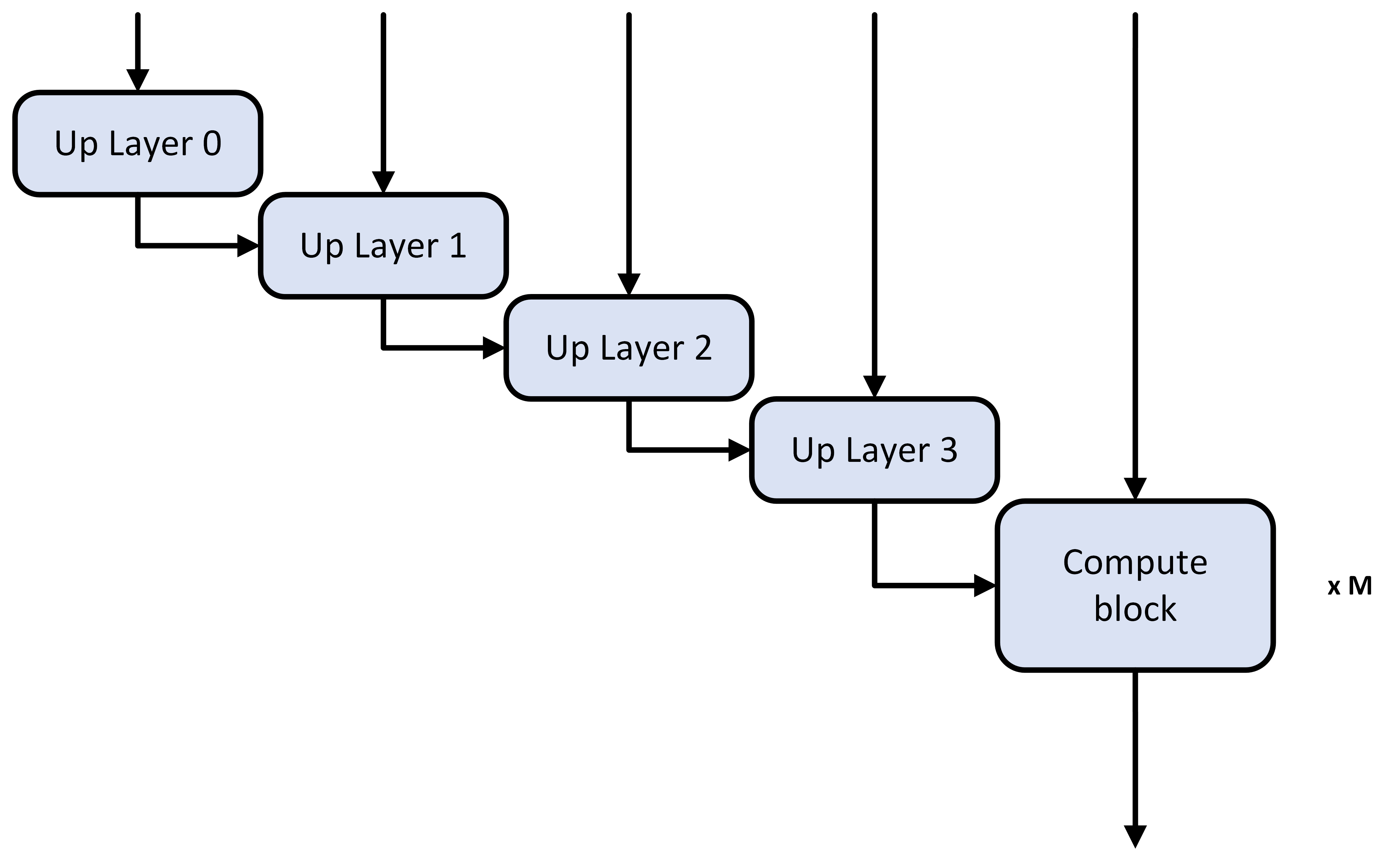}}
\caption{U-Net decoder.\label{fig:unet_decoder}}
\end{figure}

\begin{figure}[ht]
\centerline{\includegraphics[scale=0.20]{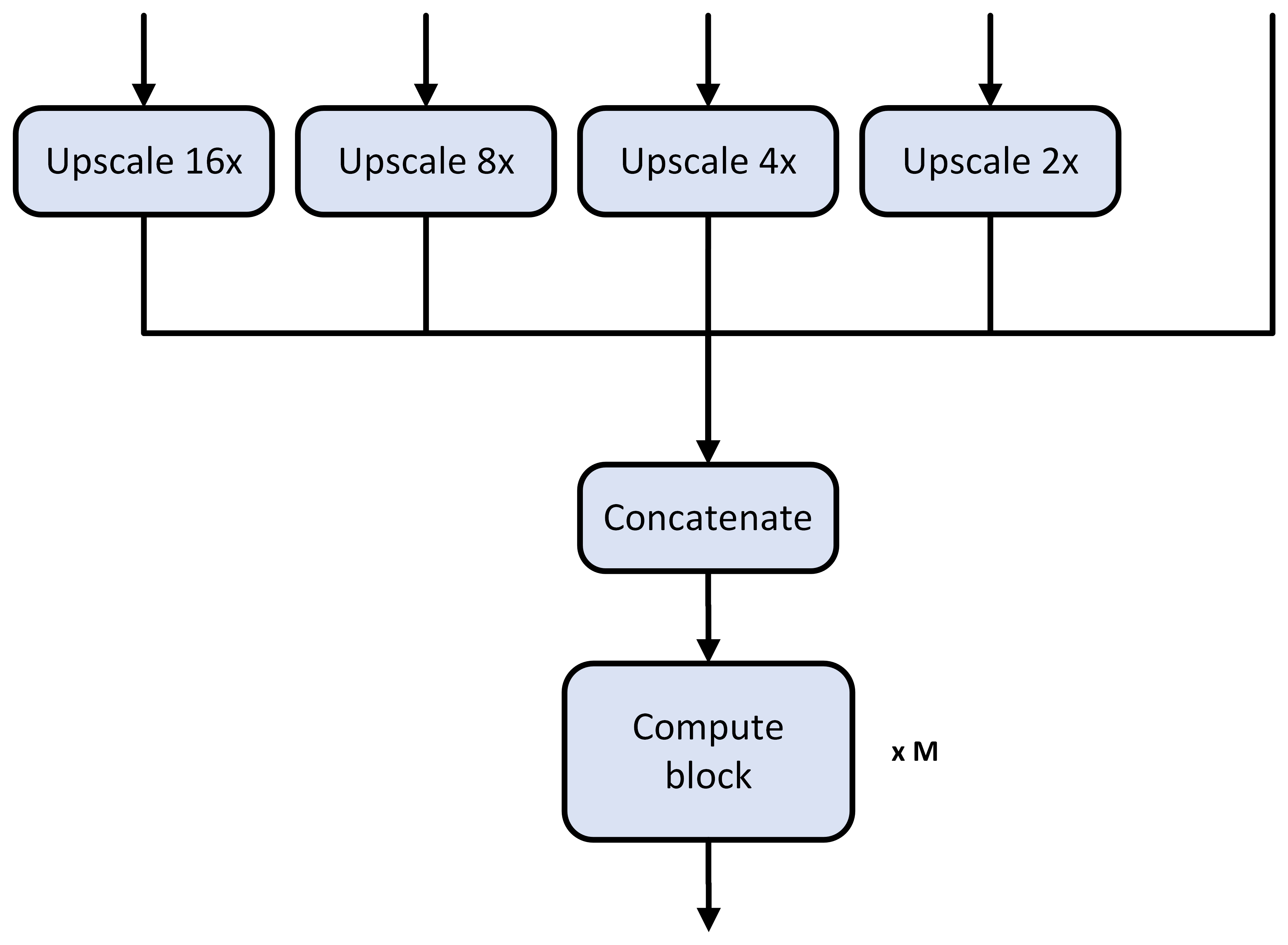}}
\caption{Half U-Net decoder.\label{fig:half_unet_decoder}}
\end{figure}

The architecture can either use the U-Net\cite{unet} decoder or the Half-U-Net\cite{half_unet} decoder variant, as shown in Figures \ref{fig:unet_decoder} and \ref{fig:half_unet_decoder}, to satisfy different requirements. The Half-U-Net architecture is a streamlined version of U-Net maintaining comparable segmentation performance to U-Net while significantly reducing parameters and computational operations.

\begin{figure}[ht]
\centerline{\includegraphics[scale=0.20]{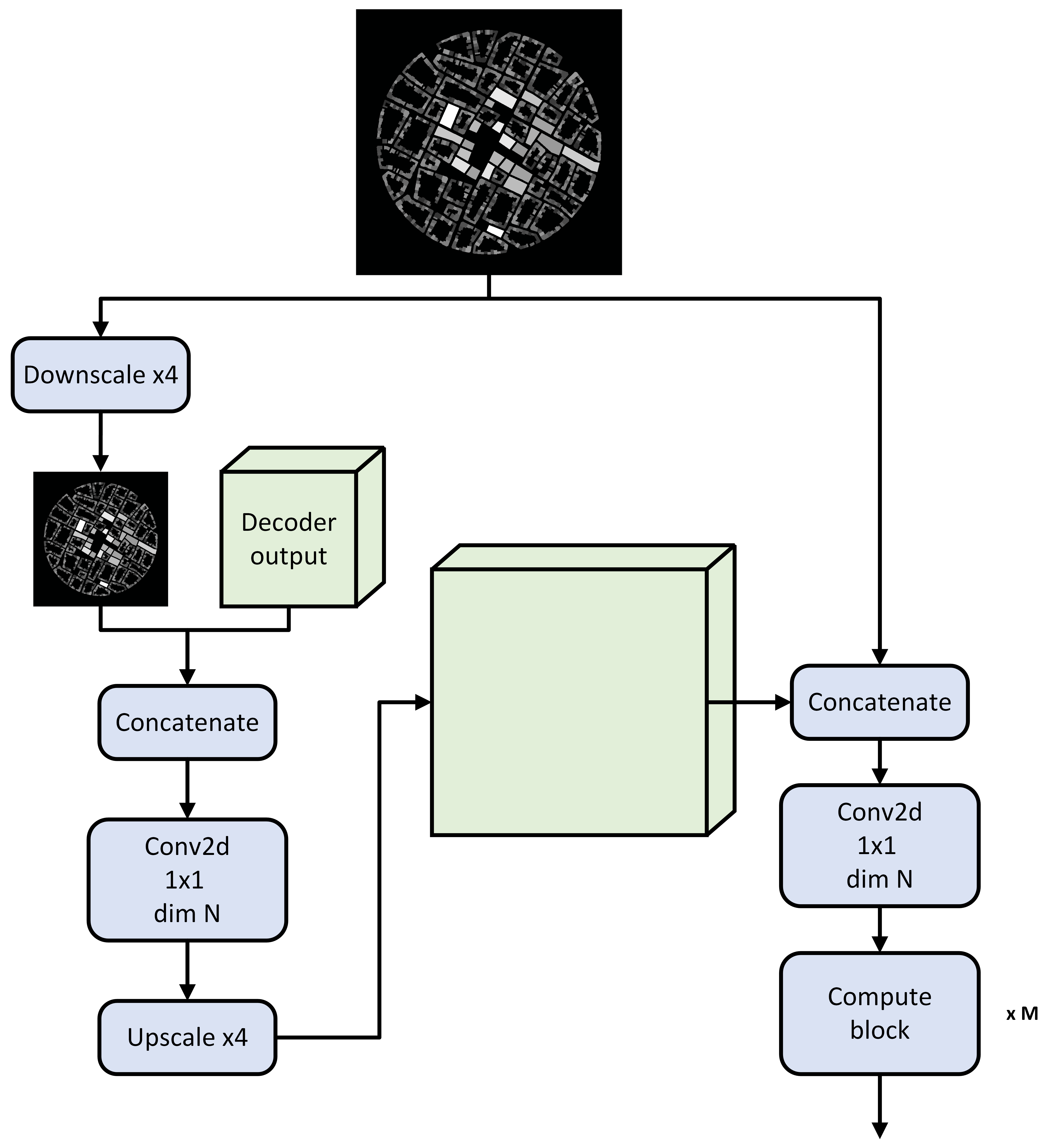}}
\caption{ResMerge block.\label{fig:res_merge}}
\end{figure}

One of the novelties of this architecture is given by the \textit{ResMerge} block shown in Figure~\ref{fig:res_merge}. Since the stem down-scales images 4x from $1024\times1024$ pixels to $256\times256$ pixels, the goal of the ResMerge module is to complement the output of the decoder with full $1024\times1024$ pixel resolution inputs.

\begin{figure}[ht]
\centerline{\includegraphics[scale=0.20]{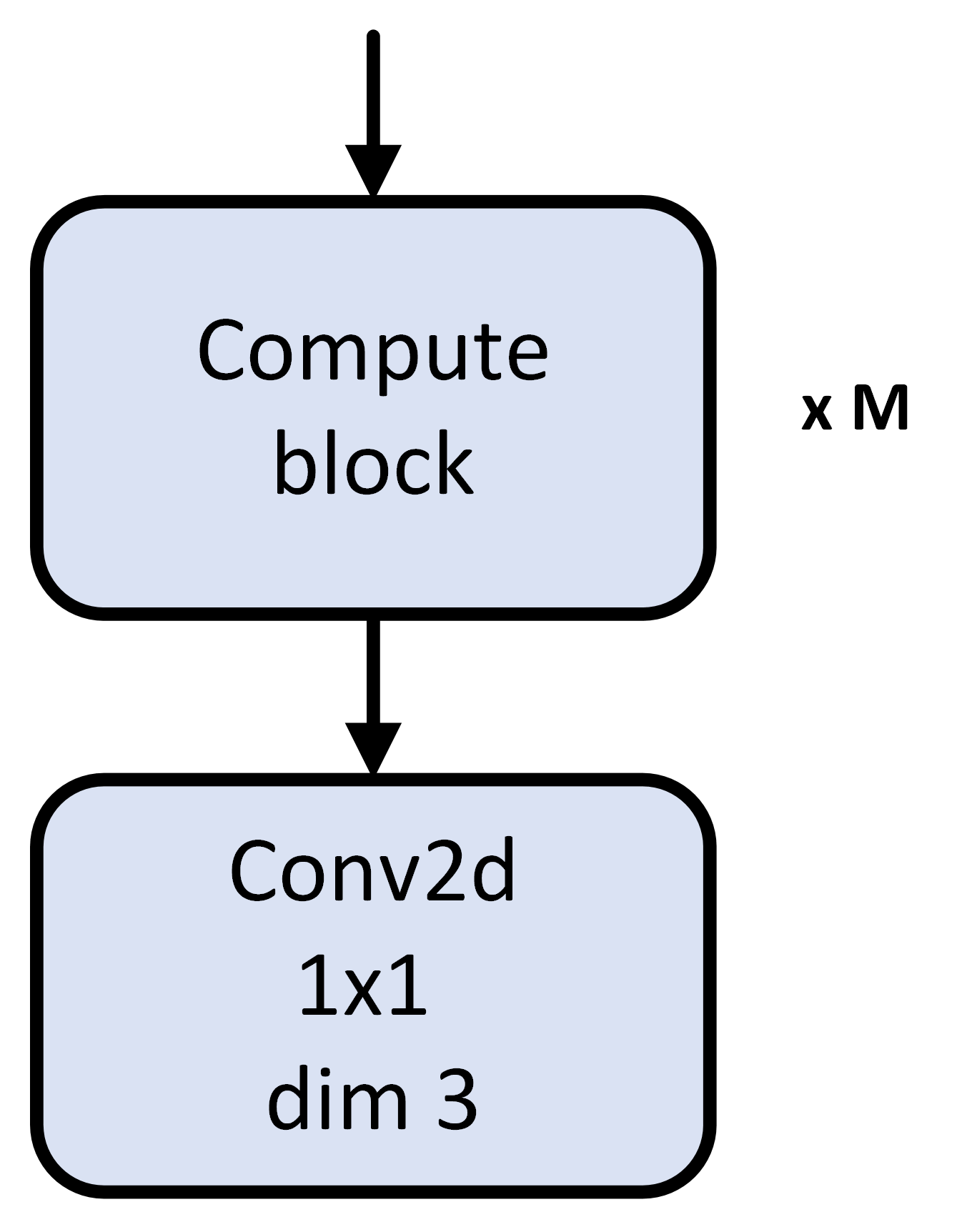}}
\caption{Output layer.\label{fig:output_layer}}
\end{figure}

Finally, the \textit{output block} serves as the terminal segment of the network, tasked with generating the final prediction (the generated image) as shown in Figure \ref{fig:output_layer}.

\begin{figure}[ht]
\centerline{\includegraphics[scale=0.20]{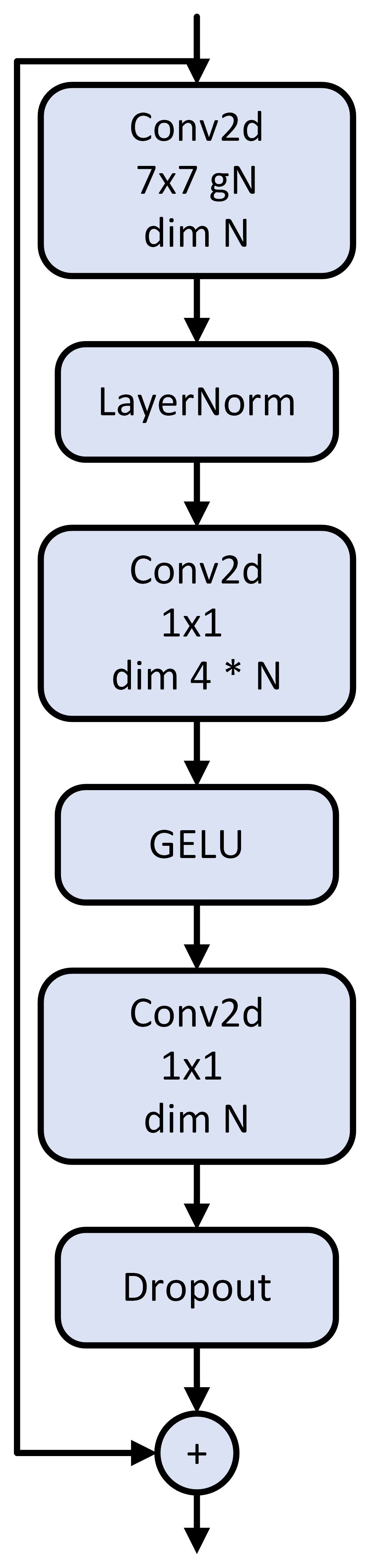}}
\caption{A ConvNext block with N output channels.\label{fig:convnext_block}}
\end{figure}

\begin{figure}[ht]
\centerline{\includegraphics[scale=0.20]{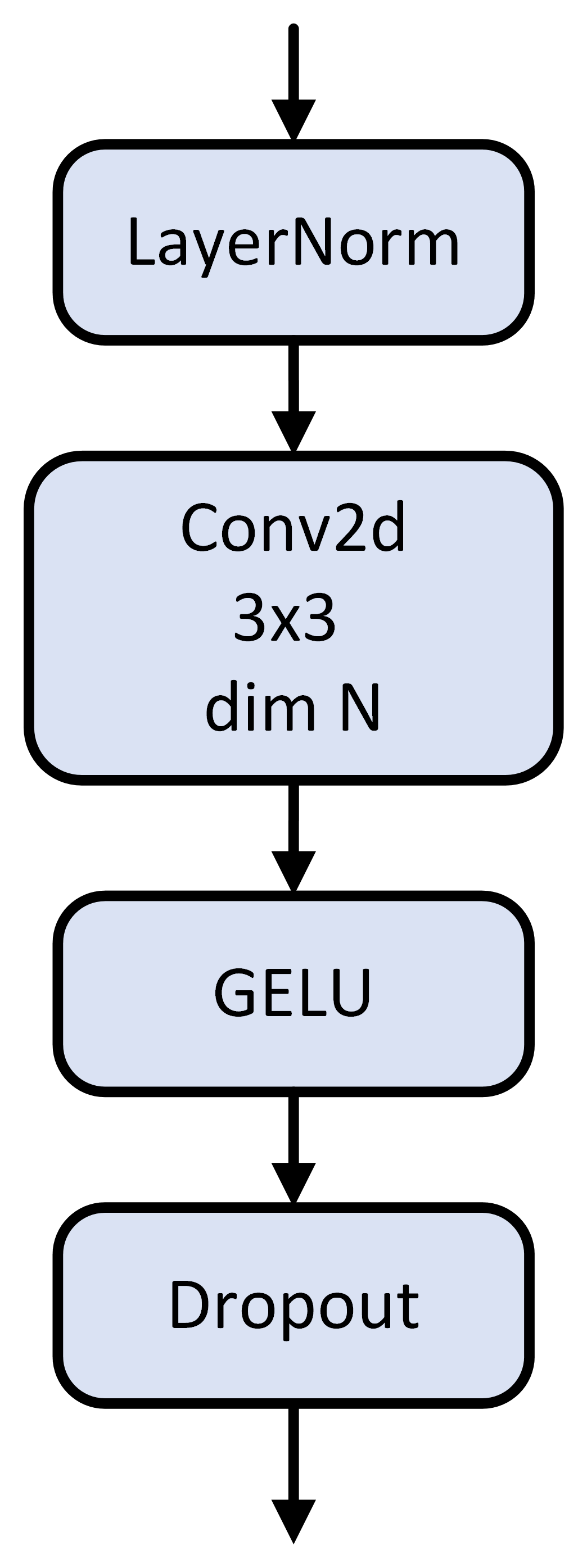}}
\caption{U-Net block with N output channels.\label{fig:unet_block}}
\end{figure}

As a summary, all configurations share the same stem module depicted in Figure~\ref{fig:stem} following the recommendation of ConvNext \cite{convnext} and the same encoder shown in Figure \ref{fig:unet_encoder} follows U-Net \cite{unet}. The encoder is then followed by one of two decoders, either a U-Net \cite{unet} decoder shown in Figure \ref{fig:unet_decoder} or a Half-U-Net \cite{half_unet} decoder shown in Figure \ref{fig:half_unet_decoder}. Finally, all configurations share a ResMerge module, depicted in Figure \ref{fig:res_merge}, and an output module seen in Figure \ref{fig:output_layer}. All blocks in the neural network are made up of one of two compute blocks, a U-Net block shown in Figure \ref{fig:unet_block}, or a ConvNext block shown in Figure \ref{fig:convnext_block}.

\section{Experimental Setting}\label{experimental_settings}
All training and experiments were conducted on a dedicated machine equipped with 4 NVIDIA RTX 3090 MSI graphics cards. The machine's storage configuration was designed to optimize performance and data management. It featured a fast system disk with a 2 TB Intel NVME SSD. For faster data processing, the machine was equipped with two 2 TB SSDs both utilizing the Linux ext4 file system. Additionally, there was a 14 TB spin disk with a Linux ext4 file system, primarily used for temporary data storage. For backup and additional temporary storage needs, a 10 TB spin disk was available also using the Linux ext4 file system. This setup ensured that the machine was well prepared for the computational demands of the training and experiments, with a clear distinction between fast data processing and storage needs. 
The code base for our experiments was developed in Python, leveraging the robust capabilities of the \textit{PyTorch} framework\cite{paszke2019pytorch}. PyTorch, known for its dynamic computational graph and intuitive design, served as the primary backbone for our deep learning models. To further enhance the training process and streamline various tasks, we incorporated the \textit{PyTorch Lightning} library\cite{Falcon_PyTorch_Lightning_2019}. This library abstracts the details of the training loop, allowing us to focus on the research code. In addition to PyTorch and PyTorch Lightning, our code relied on several other essential libraries. These included \textit{torchvision}4\cite{marcel2010torchvision} for computer vision tasks, \textit{numpy}\cite{harris2020array} for numerical operations, and \textit{matplotlib}\cite{Hunter:2007} 
for visualization and plotting. The specific versions of the primary libraries used were PyTorch Lightning (version 2.0.0 or higher) and Torch (version 2.0.0 or higher).

Four architectures are tested, combining both block types and both decoder types. The table of architectures can be seen in Table \ref{tab:architectures}. 
\begin{table}[ht]
    \centering
    \caption{Overview of the tested architectures.}
    \begin{tabular}{lll}
    \toprule
                  Name & Decoder type & Block type \\
    \midrule
    Half-U-NeXt &   Half U-Net &   ConvNeXt \\
    Half-U-Net &   Half U-Net &      U-Net \\
    U-NeXt &        U-Net &   ConvNeXt \\
    U-Net &        U-Net &      U-Net \\
    \bottomrule
    \end{tabular}
    \label{tab:architectures}
\end{table}

For each architecture a random search with 128 samples is performed, training for 30 epochs. The hyperparameter spaces for the U-Net and Half U-Net decoder architectures can be seen in Tables \ref{tab:table_unet_hyperparams} and \ref{tab:table_half_unet_hyperparams}, respectively.

\begin{table}[h]
\centering
    \caption{Tested hyperparameters for U-Net decoder}
\begin{tabularx}{\columnwidth}{lX}
        \toprule
        Hyperparameter & Values \\
        \midrule
        Encoder Channels & [32, 64, 128, 256, 512], \newline [64, 128, 256, 512, 1024], \newline [128, 256, 512, 1024, 2048] \\ 
    Decoder Channels & Encoder channels reversed \\ 
    Encoder Blocks & [1, 1, 1, 1, 1], \newline [2, 2, 2, 2, 2], \newline [4, 4, 4, 4, 4] \\ 
    Decoder Blocks & [1, 1, 1, 1, 1], \newline [2, 2, 2, 2, 2], \newline [4, 4, 4, 4, 4] \\ 
    Output Blocks & 1, 2, 4 \\ 
    ResMerge Blocks & 1, 2, 4 \\ 
    Dropout & 0.1, 0.2, 0.3 \\ 
    \end{tabularx}
    \label{tab:table_unet_hyperparams}
\end{table}

\begin{table}[]
    \centering
        \caption{Tested hyperparameters for Half U-Net decoder}
\begin{tabularx}{\columnwidth}{lX}
        \toprule
        Hyperparameter & Values \\
        \midrule
        Encoder Channels & [32, 32, 32, 32, 32], \newline [64, 64, 64, 64, 64], \newline [128, 128, 128, 128, 128] \\ 
    Decoder Channels & Same as encoder \\ 
    Encoder Blocks & [1, 1, 1, 1, 1], \newline [2, 2, 2, 2, 2], \newline [4, 4, 4, 4, 4] \\ 
    Decoder Blocks & [1, 1, 1, 1, 1], \newline [2, 2, 2, 2, 2], \newline [4, 4, 4, 4, 4] \\ 
    Output Blocks & 1, 2, 4 \\ 
    ResMerge Blocks & 1, 2, 4 \\ 
    Dropout & 0.1, 0.2, 0.3 \\ 
    \end{tabularx}
    
        \label{tab:table_half_unet_hyperparams}
    \end{table}

The training uses an AdamW optimizer \cite{adamw} with a learning rate of $0.001$ and a Huber Loss \cite{huber}.

Once the best hyperparameters for each architecture have been found, 10 instances of the best models with different random seeds are trained for each architecture.

\section{Results and Discussion}\label{results_and_discussion}

\begin{figure}[ht]
\centerline{
\includegraphics[width=\columnwidth]{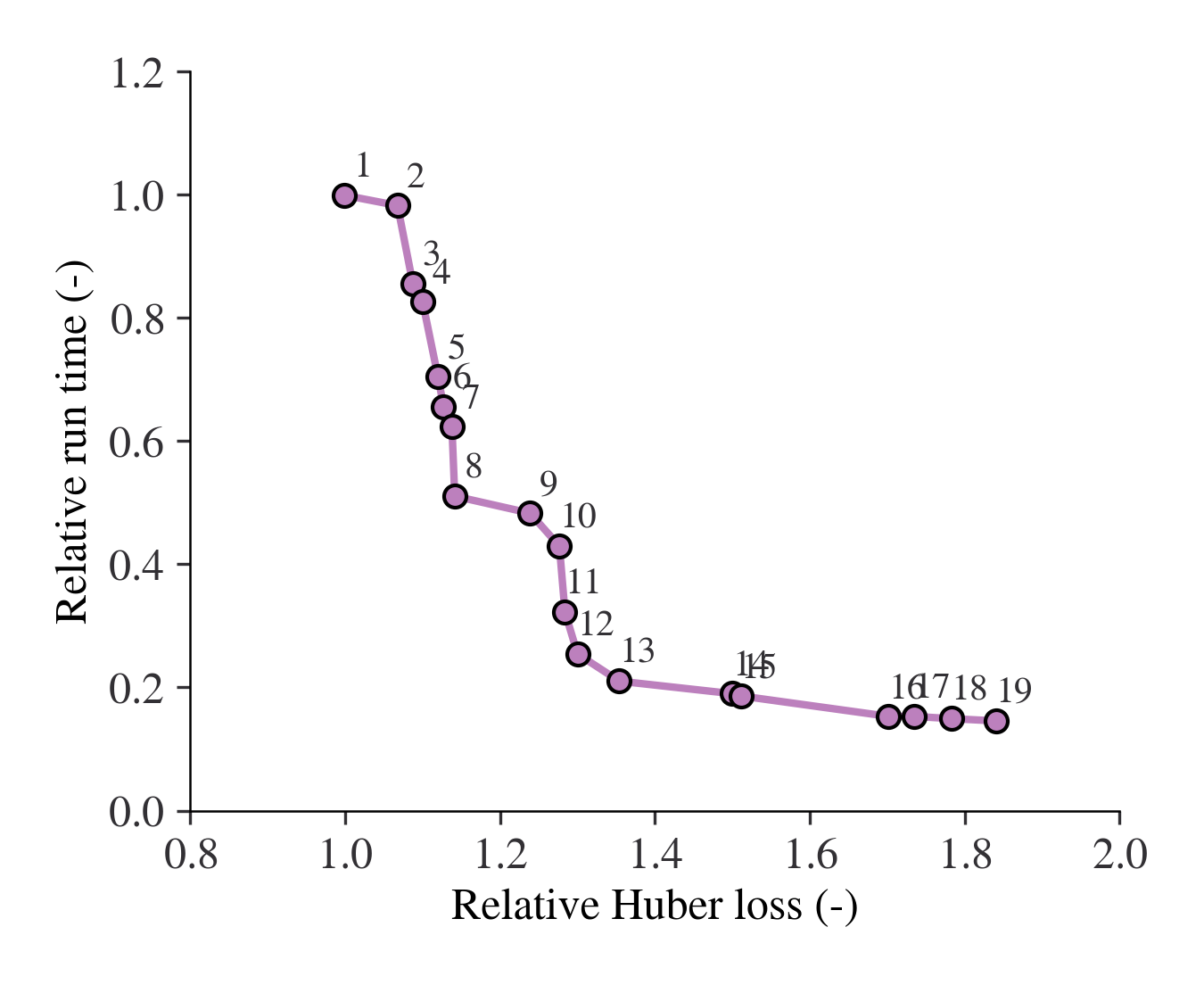}}
\caption{Pareto front showing the optimal models in terms of Huber loss for a given run-time performance point.\label{fig:pareto_front}}
\end{figure}

In the realm of multi-objective optimization, our study is anchored around two primary objectives: minimizing the Huber loss and optimizing the runtime. Such problems, by their very nature, do not yield a singular optimal solution but rather present a spectrum of them. This spectrum, termed the Pareto front, represents configurations where any incremental improvement in one objective would necessitate a compromise in the other. A total of 512 configurations were assessed in this work. These configurations were systematically designed to explore the performance implications of different block types (ConvNext and U-Net) and decoder architectures (U-Net and Half-U-Net). The aim was to discern the relative advantages and trade-offs associated with each combination, thereby providing a comprehensive understanding of their individual and collective impacts on the overarching objectives. The Pareto front for the 512 models we assessed is visually represented in Figure \ref{fig:pareto_front}. 

For a more granular understanding, Table \ref{tab:pareto_front} offers a breakdown of the performance metrics associated with each configuration. The columns of the table can be understood as follows:
\begin{itemize}
    \item \textbf{Config}: This column enumerates the unique identifier for each configuration, facilitating easy reference.
    \item \textbf{Huber Loss}: Represents the Huber loss value for each configuration, providing a measure of the model's prediction accuracy.
    \item \textbf{Runtime} (ms): Indicates the time taken, in milliseconds, for each configuration to execute, serving as a metric for computational efficiency.
    \item \textbf{Relative Loss} and \textbf{Relative Runtime}: These columns provide normalized metrics, offering a comparative perspective on the performance of each configuration against a baseline.
    \item \textbf{Block Type}: Specifies the type of block (either ConvNext or U-Net) employed in the configuration.
    \item \textbf{Decoder Type}: Denotes the decoder architecture (either U-Net or Half-U-Net) utilized in the configuration.
    \item \textbf{Parameters} (G): Provides the number of parameters, in gigabytes, associated with each configuration, giving an insight into the model's complexity.
    \item \textbf{Multiply Adds} (T): Represents the number of multiply-add operations, in terabytes, required by each configuration, offering another lens to gauge computational complexity.
\end{itemize}

\begin{table*}[ht]
    \centering
     \caption{Configurations in the pareto front of all tested configurations for all models.}
    \begin{tabular}{rrrrrllrr}
    \toprule
    Config & Huber loss & Runtime (ms) & Relative loss & Relative runtime & Block type & Decoder type & Parameters (G) & Multiply Adds (T) \\
    \midrule
         0 &     0.0090 &                 0.5400 &        1.0000 &           1.0000 &   ConvNeXt &        U-Net &       160.5219 &          349.2152 \\
         1 &     0.0096 &                 0.5304 &        1.0673 &           0.9823 &   ConvNeXt &        U-Net &        62.9737 &          146.3821 \\
         2 &     0.0098 &                 0.4621 &        1.0878 &           0.8557 &   ConvNeXt &        U-Net &        40.3998 &          124.0133 \\
         3 &     0.0099 &                 0.4462 &        1.0996 &           0.8263 &   ConvNeXt &        U-Net &       137.7440 &          260.6630 \\
         4 &     0.0101 &                 0.3804 &        1.1201 &           0.7045 &   ConvNeXt &        U-Net &        92.8223 &          216.8194 \\
         5 &     0.0101 &                 0.3539 &        1.1276 &           0.6554 &   ConvNeXt &        U-Net &        62.9537 &          125.5112 \\
         6 &     0.0102 &                 0.3358 &        1.1382 &           0.6219 &   ConvNeXt &        U-Net &        34.6369 &           90.7766 \\
         7 &     0.0103 &                 0.2753 &        1.1413 &           0.5098 &   ConvNeXt &   Half U-Net &         2.2268 &           84.1565 \\
         8 &     0.0112 &                 0.2607 &        1.2400 &           0.4827 &   ConvNeXt &   Half U-Net &         0.9639 &           44.8084 \\
         9 &     0.0115 &                 0.2310 &        1.2767 &           0.4279 &   ConvNeXt &   Half U-Net &         0.8547 &           37.6738 \\
        10 &     0.0115 &                 0.1742 &        1.2835 &           0.3226 &   ConvNeXt &   Half U-Net &         1.2582 &           53.9973 \\
        11 &     0.0117 &                 0.1376 &        1.2993 &           0.2548 &      U-Net &        U-Net &        34.6253 &           69.9619 \\
        12 &     0.0122 &                 0.1139 &        1.3536 &           0.2110 &      U-Net &        U-Net &        22.0487 &           57.8741 \\
        13 &     0.0135 &                 0.1031 &        1.5001 &           0.1909 &      U-Net &   Half U-Net &         0.4962 &           30.0155 \\
        14 &     0.0136 &                 0.1004 &        1.5115 &           0.1859 &      U-Net &        U-Net &         5.5324 &           29.9998 \\
        15 &     0.0153 &                 0.0828 &        1.7021 &           0.1533 &      U-Net &   Half U-Net &         0.3109 &           26.7919 \\
        16 &     0.0156 &                 0.0826 &        1.7348 &           0.1530 &      U-Net &   Half U-Net &         0.3109 &           26.7919 \\
        17 &     0.0160 &                 0.0800 &        1.7815 &           0.1482 &      U-Net &   Half U-Net &         0.0937 &           22.2221 \\
        18 &     0.0165 &                 0.0785 &        1.8393 &           0.1453 &      U-Net &   Half U-Net &         0.0937 &           22.2221 \\
    \bottomrule
    \end{tabular}
   
    \label{tab:pareto_front}
\end{table*}

\begin{figure}[ht]
\centerline{
\includegraphics[width=\columnwidth]{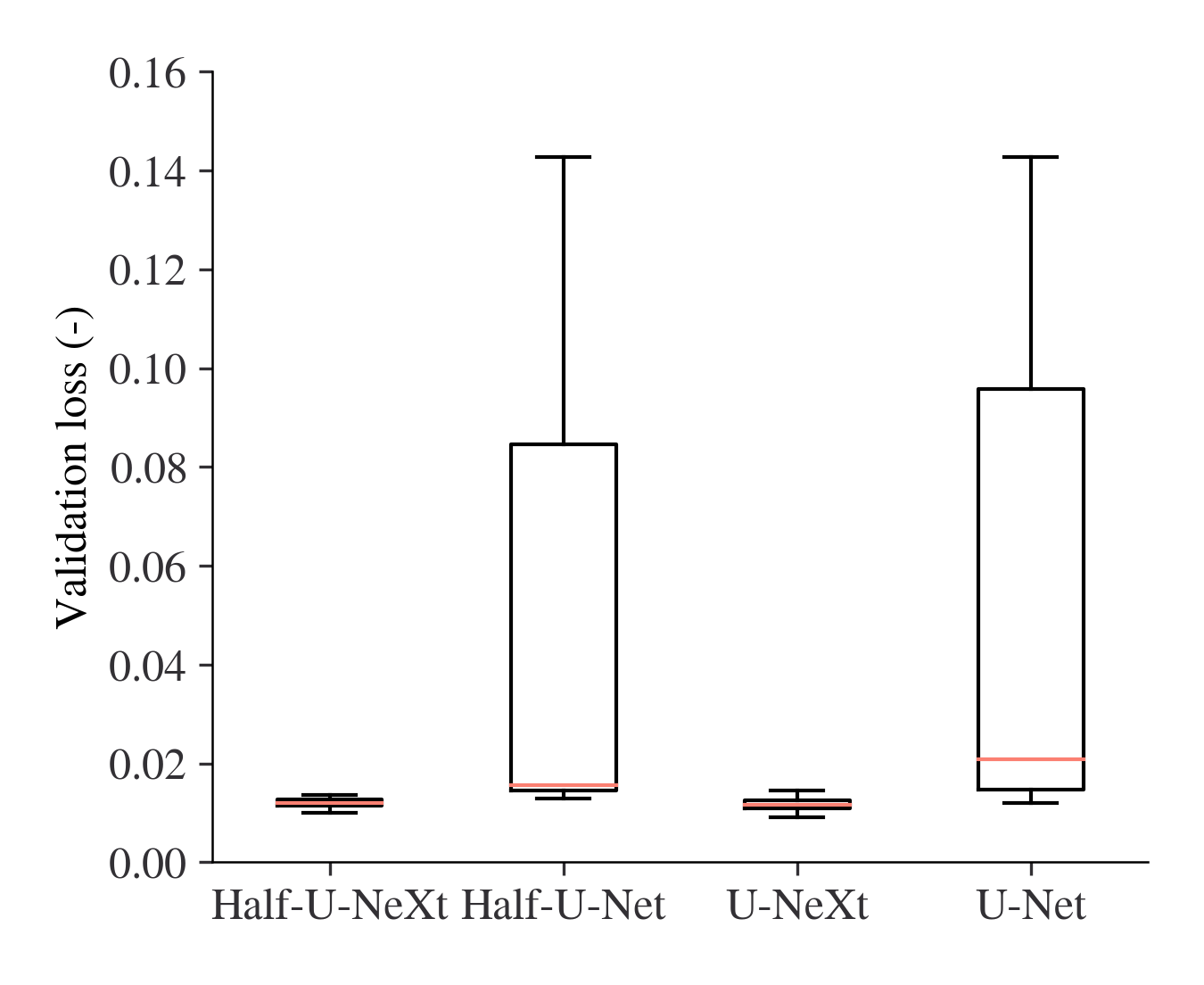}}
\caption{Comparison of the best models for each architecture in terms of spread in validation loss. Each model is trained 10 times with different random seeds.\label{fig:best_architectures}}
\end{figure}

A few salient observations can be drawn from the analysis of the Pareto front:

\begin{itemize}
    \item \textbf{ConvNext vs. U-Net Blocks}: Configurations employing the ConvNext blocks consistently demonstrate superior performance over those using the U-Net blocks as shown also in Figure \ref{fig:best_architectures}. For instance, the first six configurations in the table, all utilizing ConvNext blocks, exhibit relatively low Huber loss values while maintaining competitive runtimes. This suggests that ConvNext blocks might have inherent architectural or computational advantages that contribute to this enhanced performance.
    \item \textbf{Decoder Architectures}: In terms of decoder architectures, both U-Net and Half-U-Net seem to offer comparable efficiencies. This is evident from the similar Huber loss and runtime values they present, even when the block types differ. For example, configurations 7, 8, and 9, all employing Half-U-Net decoders, have performance metrics in the same ballpark as configurations 11, 12, and 14, which use U-Net decoders.
    \item \textbf{Computational Complexity}: A noteworthy aspect is the significant reduction in parameters and multiply adds in some configurations, especially those employing the Half-U-Net decoder. For instance, configuration 7 has only 2.2268G parameters, a stark reduction from the 160.5219G parameters of configuration 0. This suggests that certain configurations, despite their reduced complexity, can still achieve competitive performance metrics.
    \item \textbf{Relative Metrics}: Relative loss and run-time values provide a normalized perspective on the performance of each configuration. It is evident that certain configurations, despite having higher absolute loss values, still maintain competitive relative runtimes, indicating their efficiency in real-world scenarios.
\end{itemize}

In conclusion, our results underscore the nuanced interplay between architectural choices and performance in multi-objective optimization problems. This enables
practical usage of the models so that configurations
can be chosen that strikes the right balance
between accuracy and efficiency.

\begin{figure*}[ht]
\centering
\includegraphics[width=0.3\textwidth]{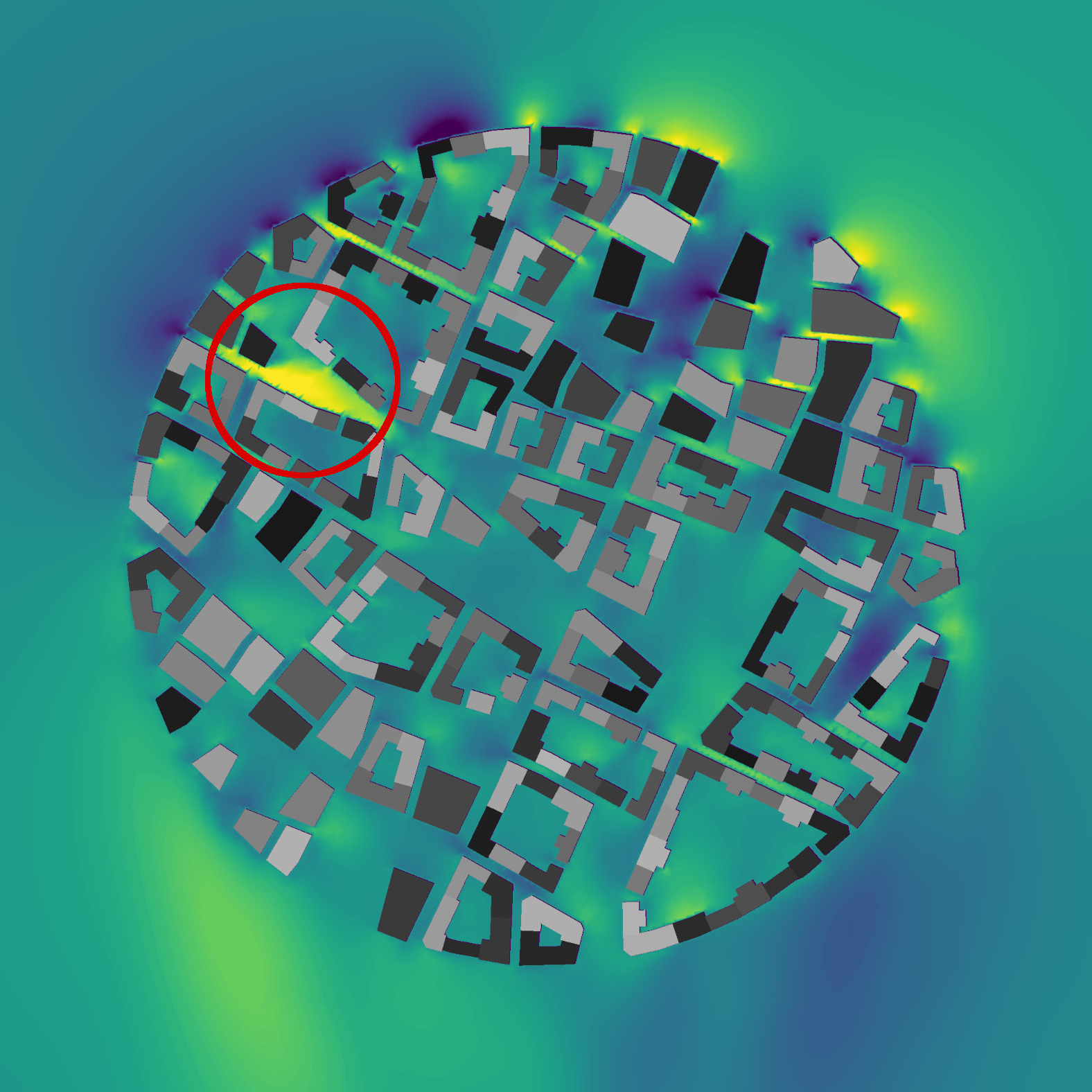}\quad%
\includegraphics[width=0.3\textwidth]{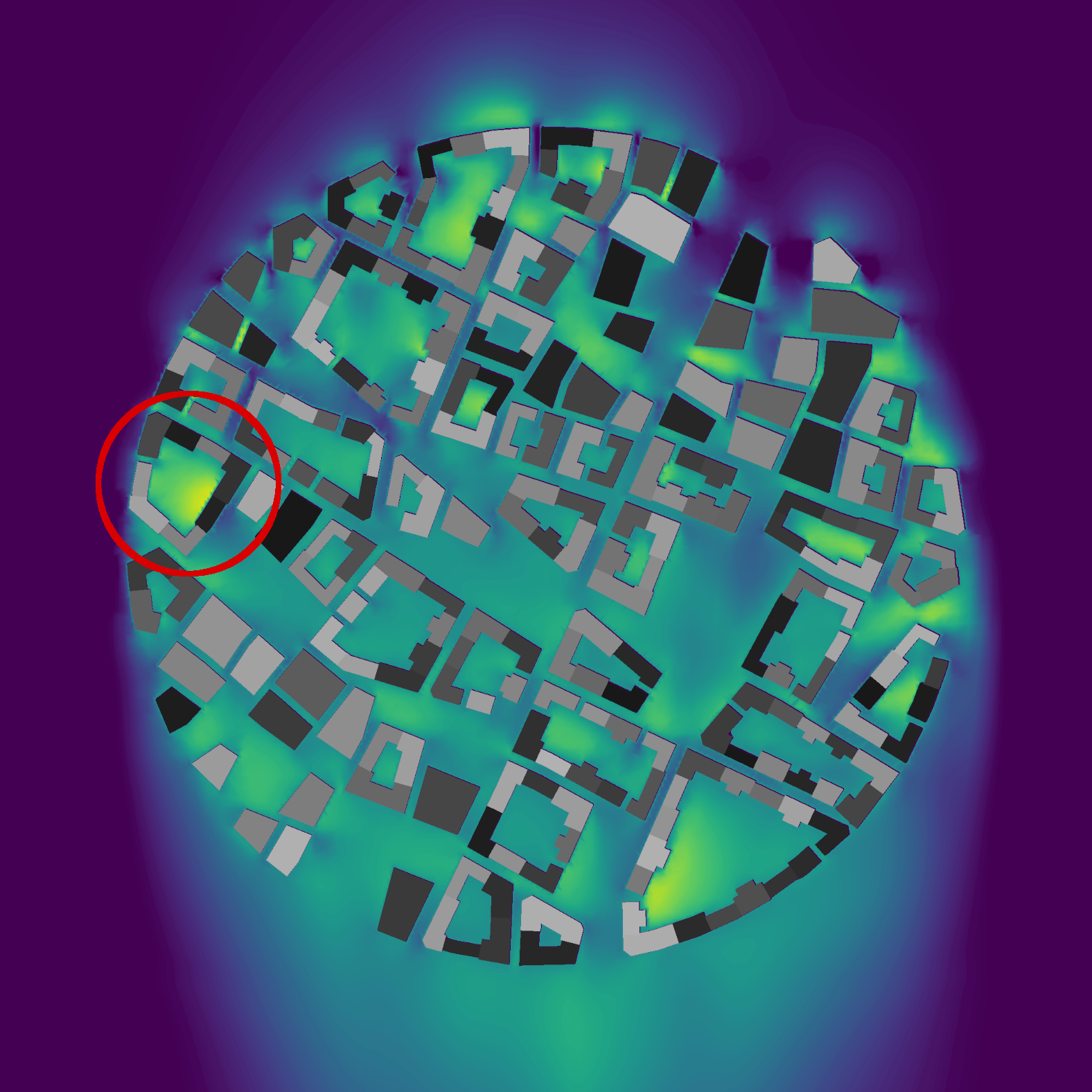}\quad%
\includegraphics[width=0.3\textwidth]{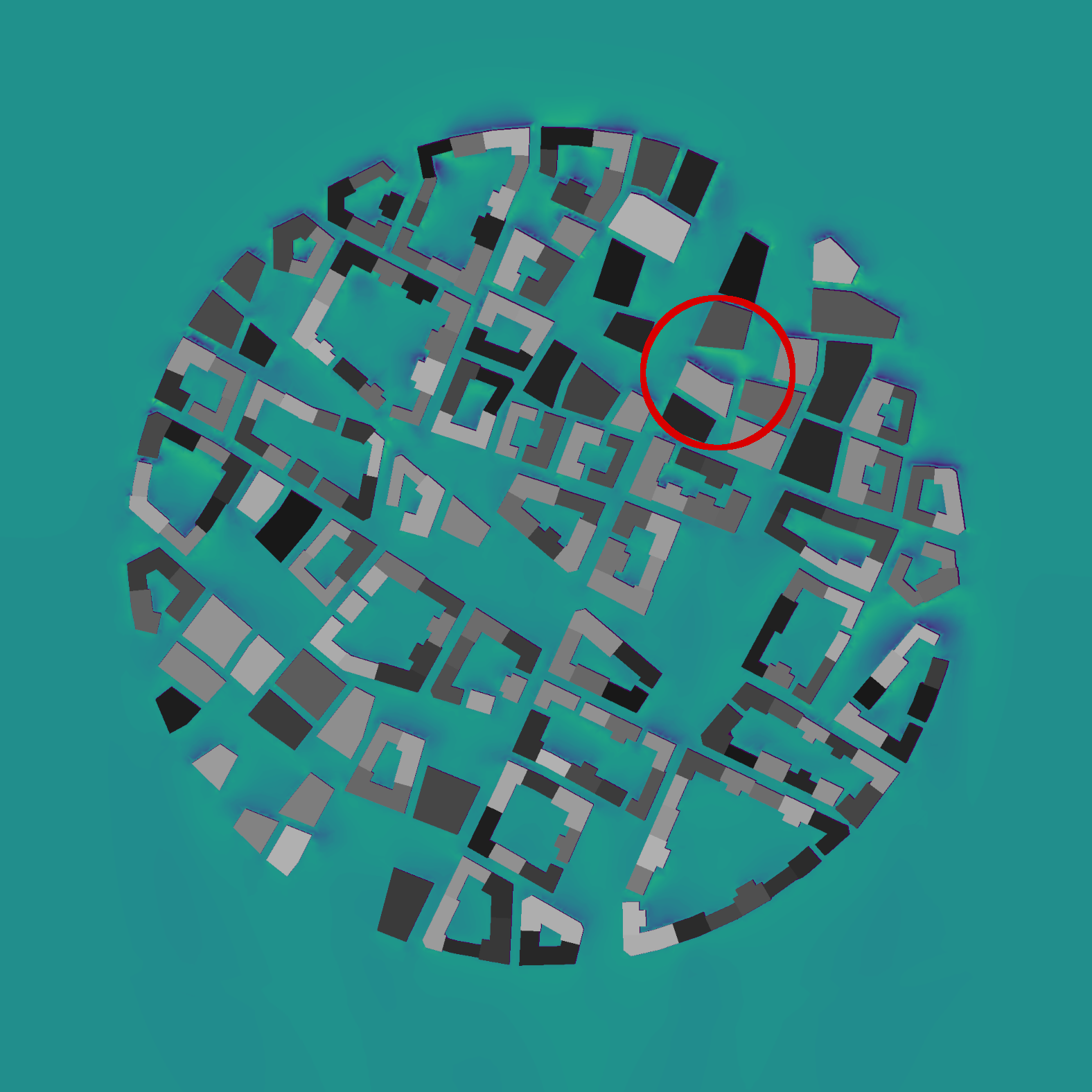}\quad%
\\\vspace{1em}
\includegraphics[width=0.3\textwidth]{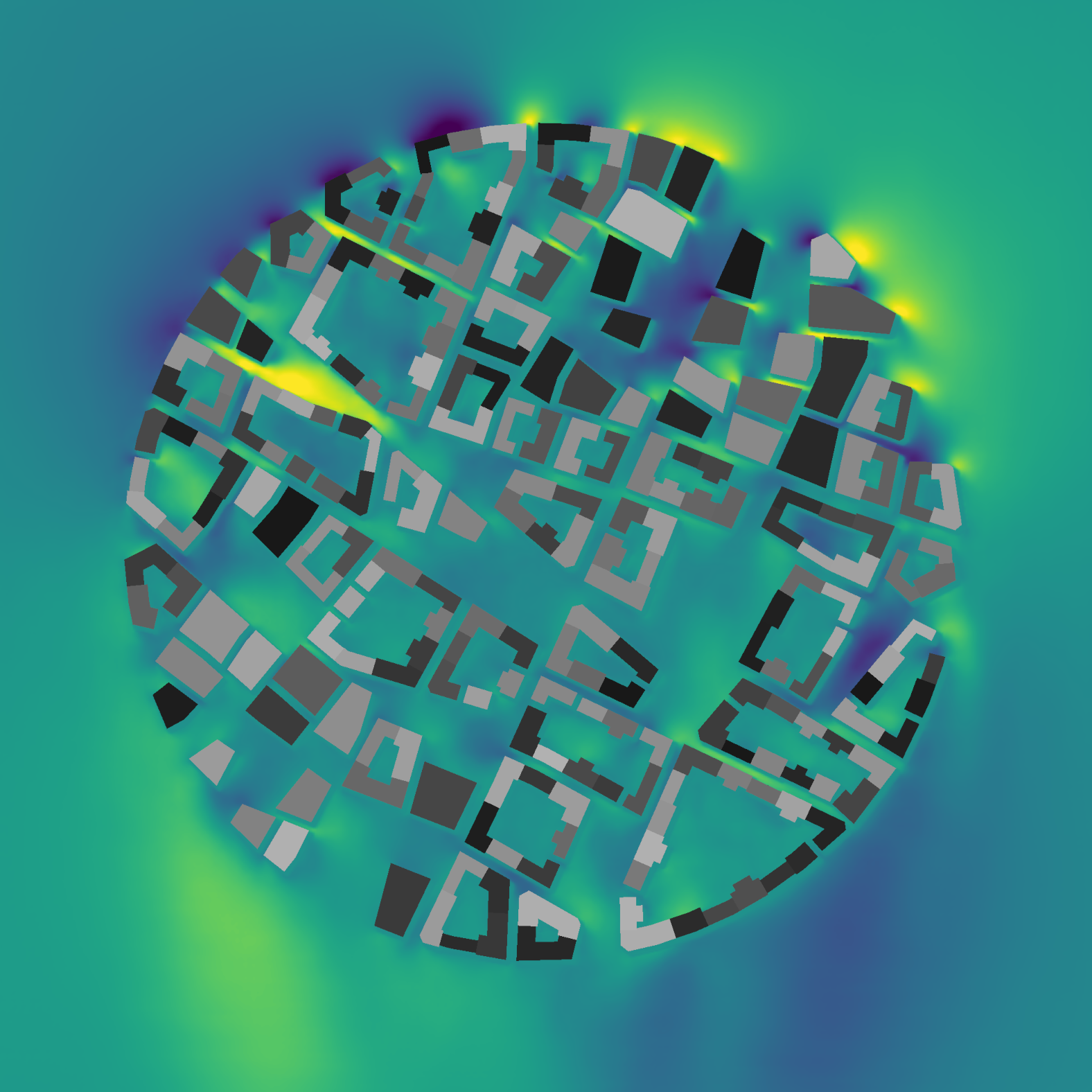}\quad%
\includegraphics[width=0.3\textwidth]{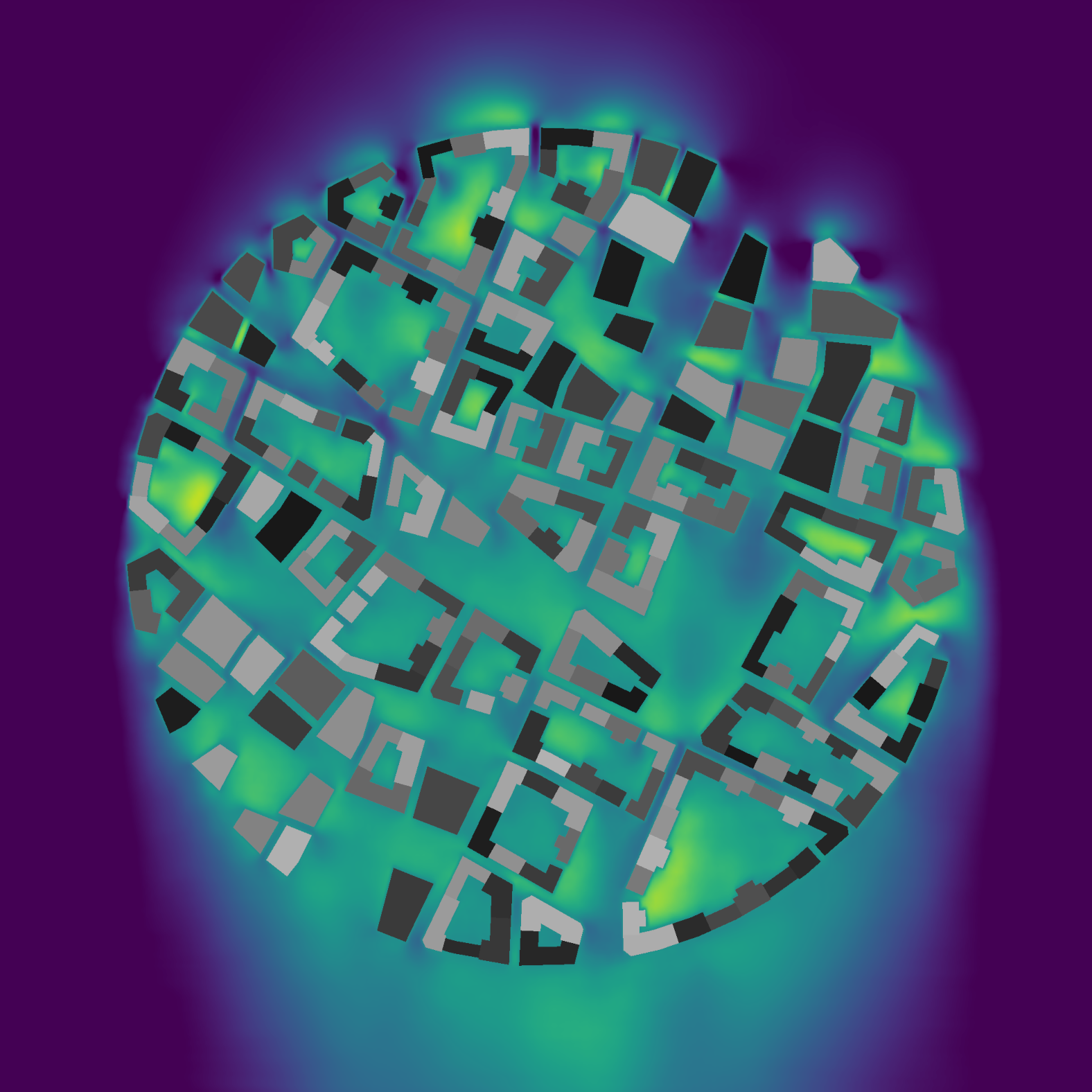}\quad%
\includegraphics[width=0.3\textwidth]{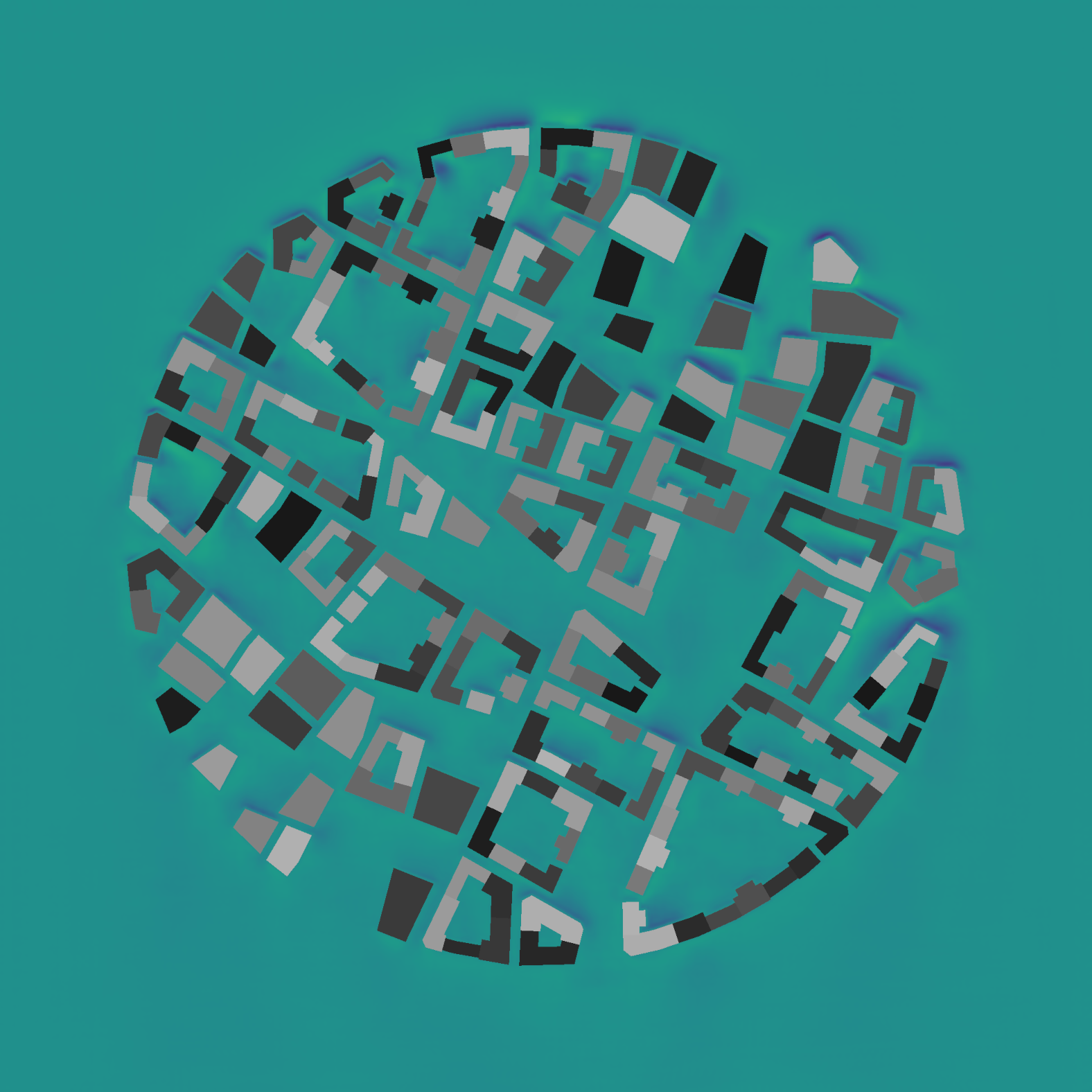}\quad%
\includegraphics[width=0.4\textwidth]{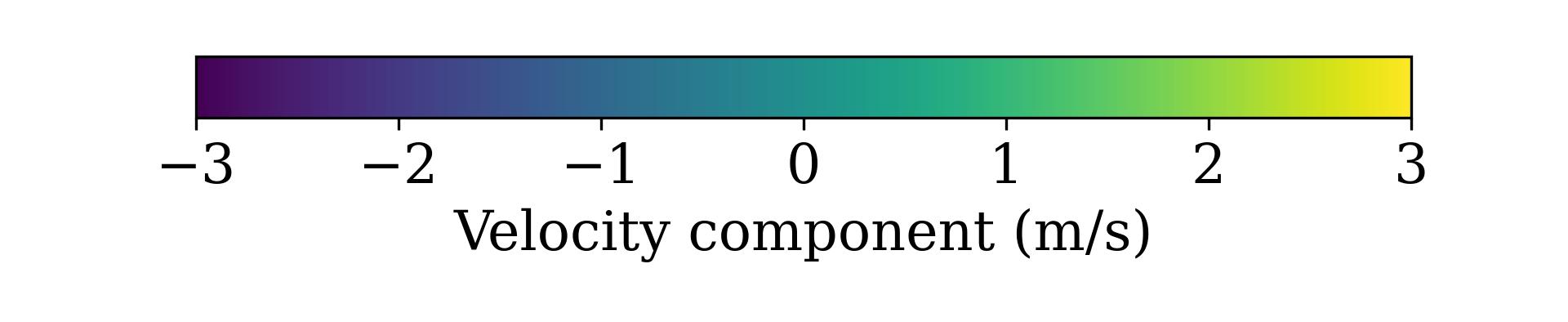}
\caption{Comparison of the three velocity vector components from the CFD simulations (top) and the model output (bottom). The geometry model is taken from the test set.
The regions highlighted by red circles are further illustrated in Figure~\ref{fig:3d_flow}.\label{fig:res_82}}
\end{figure*}

\begin{figure}[ht]
\centering
\includegraphics[width=0.5\textwidth]{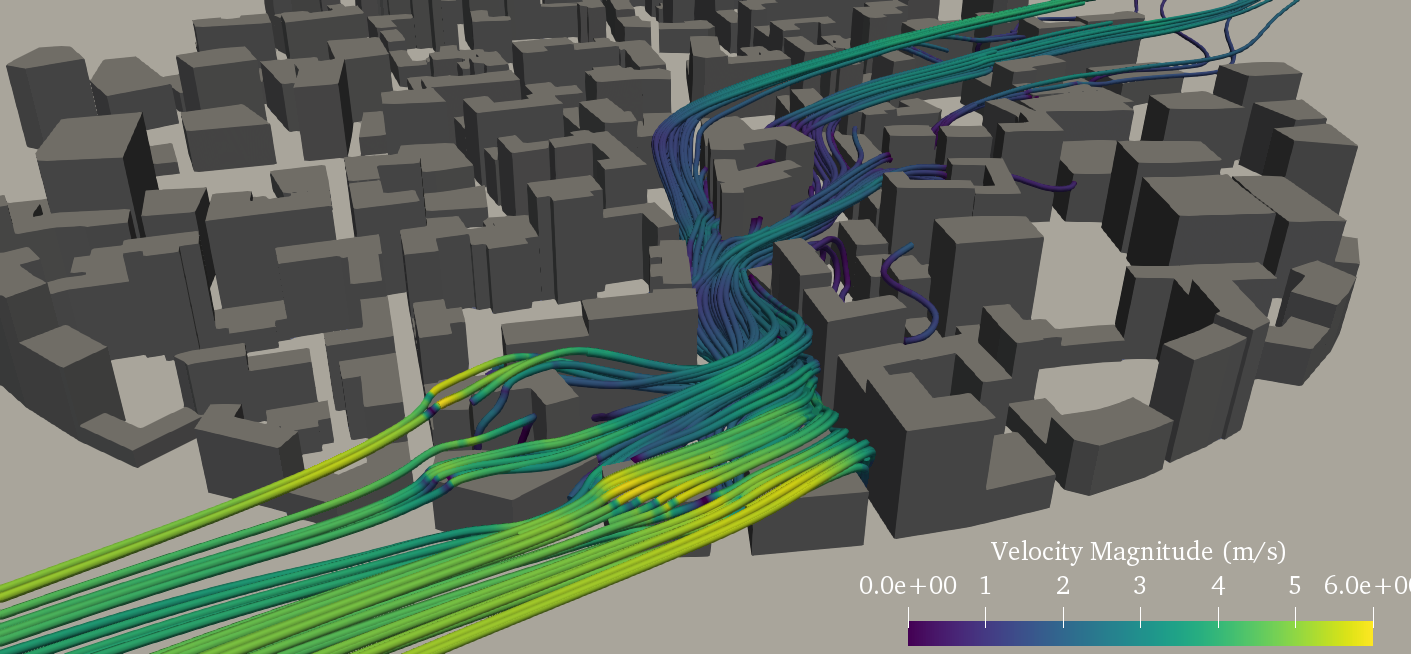}
\\\vspace{1em}
\includegraphics[width=0.5\textwidth]{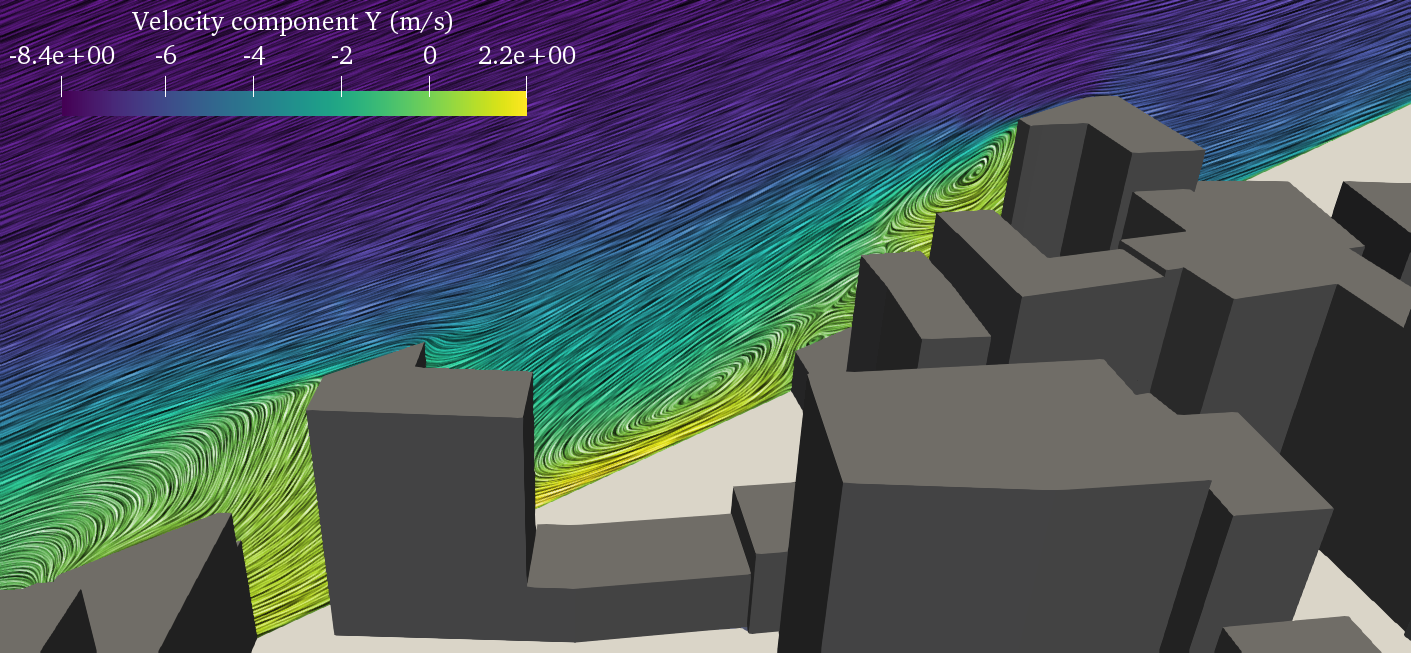}
\\\vspace{1em}
\includegraphics[width=0.5\textwidth]{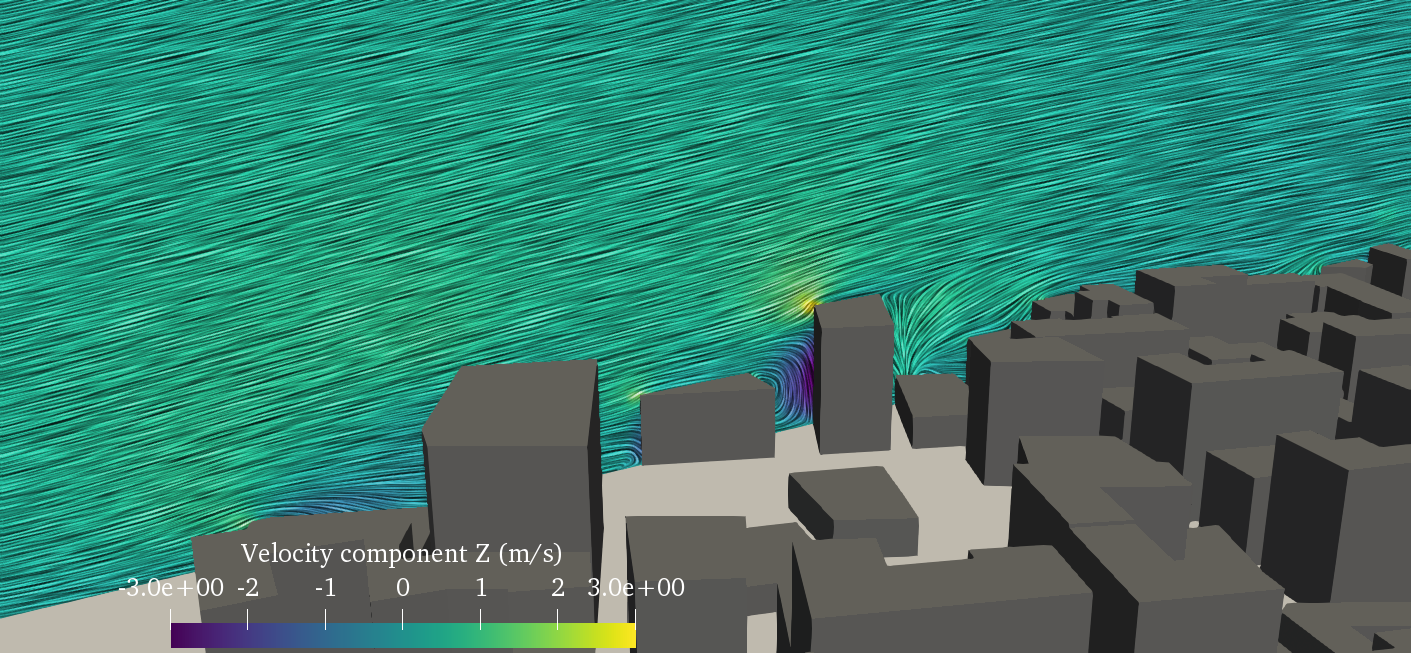}
\caption{3D flow details taken from the CFD training data, corresponding to the three circled regions in Figure~\ref{fig:res_82}.}
\label{fig:3d_flow}
\end{figure}

To further illustrate the performance of the method,
we explore one case of the test set in more detail using
configuration~7, which has a good blend of low Huber loss 
and low inference time.
Figure~\ref{fig:res_82} shows a comparison of the CFD
simulation output and the model output. The first row is the 
CFD simulation, the second row is the proposed ML model, and the three columns show the three components of the velocity vector, respectively. The wind direction is from the top to the bottom (negative $y$-direction). From a qualitative view, the agreement is excellent, as is expected from the low Huber loss.
Delving deeper into the results, we look at the flow features in the regions represented by the red circles in the CFD results. The first region shows a region with high $x$-component of velocity. The top image of Figure~\ref{fig:3d_flow} shows a 3D view of the situation, illustrated by streamlines showing how the wind hits the row of taller buildings, and being diverted towards the left into the street canyon. The street canyon gets narrower, which further accelerates the flow.
The second region shows a region where the $y$-component of the velocity goes in the opposite direction of the wind direction. The middle image of Figure~\ref{fig:3d_flow}
shows the flow pattern in this region, visualized by
line integral convolution in a vertical slice. We see that
the region consists of taller buildings with a courtyard in the middle. As the wind passes over the tall building, a recirculation zone is created in the courtyard, leading to the wind moving in the other direction at pedestrian level.
The third region shows a region where the $z$-component of the velocity has a high negative value. As further illustrated in the bottom figure of Figure~\ref{fig:3d_flow}, this region has a tall building with a lower building in front of it. As the wind hits the taller building, some of it is diverted downwards into the gap between the two buildings, leading to a high negative vertical velocity. 
These examples illustrate how a model trained on 2D data
is still capable of reproducing the velocities at pedestrian level resulting from complex 3D flow patterns in an urban environment.

\section{Discussion}\label{discussion}

The presented model shows that in the context of urban wind flow and pedestrian wind comfort, a 2D plane of the 3D flow field can be predicted using the 2D building height map as input. This enables a fast prediction, which was the objective of this work. Even the most accurate configuration analyzed in this work has an inference run-time of less than a second. This allows the use of the method in early-stage design work, where speed is more important than accuracy. There are limitations of the current methodology when it comes to usage on more complex, detailed geometries. 
The current data set does not include overhangs or skyways, which would not work with the current projection-based approach. 
Complexities in the building shapes, such as unique geometric features or slanted roofs, are also not part of the dataset. These could be included, but it would increase the number of simulations required to generate a sufficiently large training data set.
The ground is also considered completely flat in this work. The terrain could be represented with an additional height-map, and this could be incorporated into the training, but again it would increase the training data and it is challenging to make representative terrain shapes for urban cases.  
Given that early-stage design often use these simplified building shapes, and that urban environments often have low terrain changes relative to the building heights, we still believe the proposed methodology is
promising and useful for industrial use cases.

\section{Conclusion}\label{conclusion}

This work presented a configurable convolutional neural network model to predict
pedestrian-level wind velocities in an urban environment. 
The model takes a 2D height map of the building geometries as input,
and outputs the three velocity components in a 2D plane at pedestrian level. This allows a fast prediction of the pedestrian comfort to be made. To generate the training data set, CFD simulations were performed for 94 different geometries representing a wide range of urban geometries, with eight different wind directions. 
The results showed the capability of the models trained on 2D data to replicate the intricacies of 3D urban flow dynamics.
By exploring different architectures and configurations of the proposed architecture, we found the Pareto front describing the trade-off between accuracy and run-time of the model with different configurations.
All the models are below 1 ms in run-time, making them suitable for real-time usage. The fastest configuration is close to seven times faster, while having a relative loss which is 1.8 times higher than the most accurate configuration. This method allows a configuration to be chosen based on the desired trade-off between inference speed and accuracy requirements.

\bmsection*{Acknowledgments}
This work was supported by the Research Council of Norway through project number 327897, led by Nablaflow AS, with SINTEF AS, University of Stavanger and Norwegian University of Science and Technology as project partners. The authors acknowledge the support of Franz Forsberg and Spacio AS in preparing urban environments for the training data.

\bibliography{wileyNJD-AMA}

\end{document}